\documentclass[fleqn,usenatbib]{mnras}

\usepackage{newtxtext,newtxmath}

\usepackage[T1]{fontenc}
\usepackage{ae,aecompl}

\usepackage{natbib}
\usepackage{journal_names}

\usepackage{graphicx}	
\usepackage{graphics}
\DeclareGraphicsExtensions{.eps,.png, .ps}
\usepackage{epsfig,epsf}
\usepackage{amsmath}	
\usepackage{amssymb}	
\usepackage{booktabs}

\usepackage{color}
\usepackage{longtable}
\newcommand{\dg}{^{\circ}}

\usepackage{dirtytalk}                               
\usepackage[nolist,nohyperlinks]{acronym}
\usepackage[normalem]{ulem}
\usepackage{hyperref}                                

\usepackage{subfigure}

\usepackage{soul}

\title[A spectroscopic analysis of EC21178-5417]{A spectroscopic analysis of the eclipsing nova-like EC21178-5417 - discovery of spiral density structures.}
\author[Z. N. Khangale et al. ]{Z. N. Khangale,$^{1,2}$\thanks{E-mail: khangalezn@saao.ac.za} P. A. Woudt,$^{1}$ S. B. Potter,$^{2}$ B. Warner,$^{1}$ D. Kilkenny,$^{3}$ \newauthor
and K. van der Heyden$^{1}$ \\
$^{1}$Department of Astronomy, University of Cape Town, Private Bag X3, Rondebosch 7701, South Africa\\
$^{2}$South African Astronomical Observatory, PO Box 9, Observatory 7935, Cape Town\\
$^{3}$Department of Physics \& Astronomy, University of the Western Cape, Private Bag X17, Bellville 7535, South Africa\\
}

\date{Accepted XXX. Received YYY; in original form \today}

\pubyear{2019}

\begin{document}
\label{firstpage}
\pagerange{\pageref{firstpage}--\pageref{lastpage}}
\maketitle


\begin{abstract}
We present phase-resolved optical spectroscopy of the eclipsing nova-like cataclysmic variable EC21178-5417 obtained between 2002 and 2013. The average spectrum of EC21178-5417 shows broad double-peaked emission lines from HeII 4686 \AA{} (strongest feature) and the Balmer series. The high-excitation feature, CIII/NIII at 4640-4650 \AA{}, is also present and appears broad in emission. 
A number of other lines, mostly HeI, are clearly present in absorption and/or emission. The average spectrum of EC21178-5417 taken at different months and years shows variability in spectral features, especially in the Balmer lines beyond H$\gamma$, from pure line emission, mixed line absorption and emission to pure absorption lines. 
Doppler maps of the HeII 4686 \AA{} emission reveal the presence of a highly-inclined asymmetric accretion disc and a two spiral arm-like structure, whereas that of the Balmer lines (H$\alpha$ and H$\beta$) reveal a more circular accretion disc. There is no evidence of a bright spot in the Doppler maps of EC21178-5417 and no emission from the secondary star is seen in the tomograms of the HeII 4686 \AA{} and Balmer lines. Generally, the emission in EC21178-5417 is dominated by emission from the accretion disc. 
We conclude that EC21178-5417 is a member of the RW Tri or UX UMa sub-type of nova-like variables based on these results and because it shows different spectral characteristics at different dates. This spectral behaviour suggests that EC21178-5417 undergoes distinct variations in mass transfer rate on the observed time scales of months and years.
\end{abstract}

\begin{keywords}
accretion, accretion discs -- binaries: close -- stars: individual: EC21178-5417 novae, cataclysmic variables
\end{keywords}

\section{Introduction}

Cataclysmic variables (CVs) are interacting binary systems consisting of an electron degenerate star (a white dwarf primary) and a low-mass secondary star. The secondary star is usually on or near the main 
sequence, rich in hydrogen, and transfers material onto the white dwarf (WD) via a gas stream and accretion disc. In addition, a bright spot is formed at the intersection of the disc and gas stream. Non-magnetic CVs are classified into two main types, namely dwarf novae and nova-like. The former, dwarf novae, show outbursts of 2-8 mag, and are subdivided into U Gem, Z Cam and SU UMa types \citep{1995CAS....28.....W}. 
The latter, nova-likes, resemble non-eruptive nova-remnants and are probably pre-novae, or remnants of prehistoric novae. Thus nova-likes show no outbursts, but a few of them show dwarf nova outbursts and some show low and high states, e.g. DW UMa \citep{1994MNRAS.266..859D,2013MNRAS.428.3559D}. 
   
Nova-like (NL) variables are disc-dominated CVs characterized by higher mass transfer rates exceeding the expected rates based on standard magnetic braking as angular momentum loss mechanism \citep{2009ApJ...693.1007T}. NLs show nearly constant accretion luminosity, and are, unlike most other CVs, almost always observed in the outburst state \citep{2010ApJ...719.1932N}. 
Thus NLs are like dwarf nova (DN) in permanent outburst and have orbital periods just above the period gap in the orbital period distribution diagram of non-magnetic CVs (e.g. Fig. 18 of \citealt{2011ApJS..194...28K}), where magnetic braking is thought to reproduce the largest mass transfer rates in the long-term evolution of CVs \citep{2001ApJ...550..897H}. 
NLs show short timescale spectroscopic and photometric variability that is similar to those of DN between outbursts. Optical spectra of NL show a wide range of distinctive features which depend on a variety of conditions, such as the temperature of the WD, inclination angle, and mass transfer from the secondary ($\dot{M}_2$). 
Based on their spectroscopic and photometric behaviour, NLs are classically divided into four distinct sub-types: UX Ursae Majoris (UX UMa), RW Triangulum (RW Tri), SW Sextantis (SW Sex), and VY Sculptoris (VY Scl). \citet{1995CAS....28.....W} gives a comprehensive review of NLs.

\smallskip
EC21178-5417 is a 13.7 magnitude eclipsing NL variable which was discovered in the 
Edinburgh-Cape Survey of blue objects \citep{1997MNRAS.287..848S}. Its orbital period is $\sim$3.708 h 
\citep{2003MNRAS.344.1193W}. This orbital period was confirmed by \cite{2008.....Z} from high-speed photometric 
data, where he also analysed spectroscopic data obtained using the Southern African Large Telescope (SALT, \citealt{2006SPIE.6269E..0AB}) in 2006 
August 17 during the primary eclipse. The average spectrum of EC21178-5417 was found to reveal strong, broad, and 
double-peaked emission from HeII 4686 \AA{} and Balmer (H$\alpha$, H$\beta$, and H$\delta$), and other HeI lines. \citet{2003MNRAS.344.1193W} found EC21178-5417 to be a rich source of dwarf nova oscillations (DNOs) and quasi-periodic oscillations (QPOs). In addition to that, they found EC21178-5417 to show other types of periodic oscillations called longer period dwarf nova oscillations (lpDNOs).  
Recent photometry studies of EC21178-54 by \cite{2017NewA...56...60B} revealed no QPOs or DNOs reported by \citet{2003MNRAS.344.1193W}. However, the observations of \cite{2017NewA...56...60B} were taken at lower signal to noise ratio. 
Also,  \citet{1972MNRAS.159..429W} and \citet{1998ApJ...499..414K} noted that for NLs, the DNOs were not always visible in the Fourier Transforms (e.g. UX UMa). \citet{2003MNRAS.344.1193W} estimated the success rate (at least once in a run) for detecting DNOs at 83 percent for EC21178-5417. 

\bigskip
In this paper we present an updated ephemeris of EC21178-5417 based on our full photometric archive of this object, spanning six epochs between 2002 and 2016 (Section \ref{sec:photometry}). 
We then present in Section \ref{sec:SpectraAnalysis} a detailed spectroscopic analysis of the original identification spectra (Edinburgh-Cape Survey), and follow-up spectra obtained in 2011 and 2013. The latter were obtained during an extended photometric campaign. 
In Section \ref{sec:Doppler}, Doppler tomography of the 2011 and 2013 spectroscopy is presented. This is followed by a brief discussion based on the results in Section \ref{sec:Discussion}.

\section{Observations} 

\subsection{Photometry}

\begin{figure}
    \centering
    \includegraphics[width=0.45\textwidth]{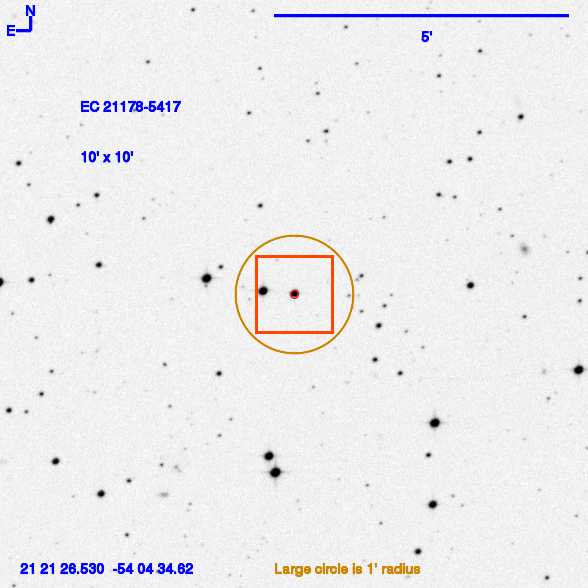}
    \caption{Digital Sky Survey image of the field of view around EC 21176-5417. The field of view of the SHOC camera (1.29' $\times$ 1.29' for the SAAO 1.9-m) is indicated with the red square and our target centre of the square and the comparison star is on the left. The orange circle marks 1' radius circle around the target star. }
    \label{figure:finder}
\end{figure}

Photometric observations of EC21178-5417 were obtained by our group from 2002 to 2016 at the South African Astronomical Observatory (SAAO) station in Sutherland using the SAAO 1.9-m Radcliffe reflector telescope. 
The Finding chart of EC21178-5417 is shown in Fig. \ref{figure:finder}. 

\subsubsection{UCT CCD photometer}

EC21178-5417 was first observed with the SAAO 1.9-m telescope using the UCT CCD photometer (University of Cape Town CCD photometer; \citealt{1995BaltA...4..519O}) between 2002 September 7 and 2002 November 30. 
The log of these observations is given in Table 1 of \cite{2003MNRAS.344.1193W}. Some of these observations were short runs, not covering the eclipse, so they were not useful in improving the ephemeris. 
Therefore, they were not used in this paper, only those dates that appears in Table \ref{tab:med-eclipse} are used. The UCT CCD photometer was operated in its white light mode in order to capture as many photons as possible. For all these observations, the UCT CCD was used in frame-transfer mode. The integration time of 5, 6 and 10 seconds were used. 

\begin{table*}
\caption{Photometric observations log.}
\label{tab:photo_obs}
\begin{center}
\begin{tabular}{lccccccc} \hline
Run No & Date & HJD of first obs & Length & $t_{in}$ & Filter & Telescope & Instrument \\ 
      & (start of night)& (+245 0000.0) & (h) & (s)   & used & used  & used \\ \hline 
S7651 & 2006 Aug 16 &3964.28141 & 5.35 & 6 & $V$ & SAAO 1.9-m  & UCT CCD \\
S7655 & 2006 Aug 17 &3965.32565 & 6.26 & 6 & $V$ & SAAO 1.9-m  & UCT CCD \\
S7659 & 2006 Aug 18 &3966.31921 & 4.36 & 6 & $V$ & SAAO 1.9-m  & UCT CCD \\ 
S7661 & 2006 Aug 19 &3967.40985 & 0.74 & 6 & $V$ & SAAO 1.9-m  & UCT CCD \\
S8097 & 2011 Oct 5  &5840.27376 & 4.12 & 0.5 & $WL$ & SAAO 1.9-m & SHOC  \\ 
\hline
\end{tabular}
\end{center}    
Notes: HJD is the Heliocentric Julian Date, $t_{in}$ is the integration time, $V$ is the Johnson $V$ filter and $WL$ is white light or unfiltered mode.
\end{table*}

The data was reduced following the recipes of \cite{1995BaltA...4..519O} and the resulting magnitudes calibrations and extinctions corrections are only approximations. The magnitude derived in this case are approximate scale and were derived with the use of hot white dwarfs standards \cite{1992AJ....104..340L}. 
Therefore, our magnitudes approximate to a $V$ scale to $\sim$0.1 mag. It is worth noting that the average mean out of eclipse magnitude ($V$) of this target ranged between 13.6 and 13.8. A total of nine eclipses resulted from these observations. 

EC21178-5417 was observed again with the SAAO 1.9-m telescope using the UCT CCD photometer  \citep{1995BaltA...4..519O} over four consecutive nights in 2006 August 16--19. Full description of these observations is given in \citet{2008.....Z}\footnote{For more details see \url{https://open.uct.ac.za/handle/11427/4412}}. We have reproduced the log of observations from \citet{2008.....Z} and is given in Table \ref{tab:photo_obs}. 
The UCT CCD photometer was equipped with a Johnson V filter and the detectors were used in frame-transfer mode. An integration time of 6 s was selected. 
These observations included at least one complete eclipse per run. We have re-reduced the 2006 data following the procedures described in \cite{1995BaltA...4..519O}. Since these observations were filtered, we calibrated our observations to the Sloan $r$ photometric system (see \citet{2014MNRAS.437..510C} for more details). 
We compared the calibration offset using hot white dwarfs with those obtained using SDSS photometry of the comparison star. It was found that there was a stable zero-point offset of 0.12$\pm$0.05 mag between $V$ and SDSS $r$. A total of seven eclipses were obtained from the 2006 data.

\subsubsection{SHOC observations}

\begin{figure}
\centering
\includegraphics[height = 5.50cm, width = 8.0cm]{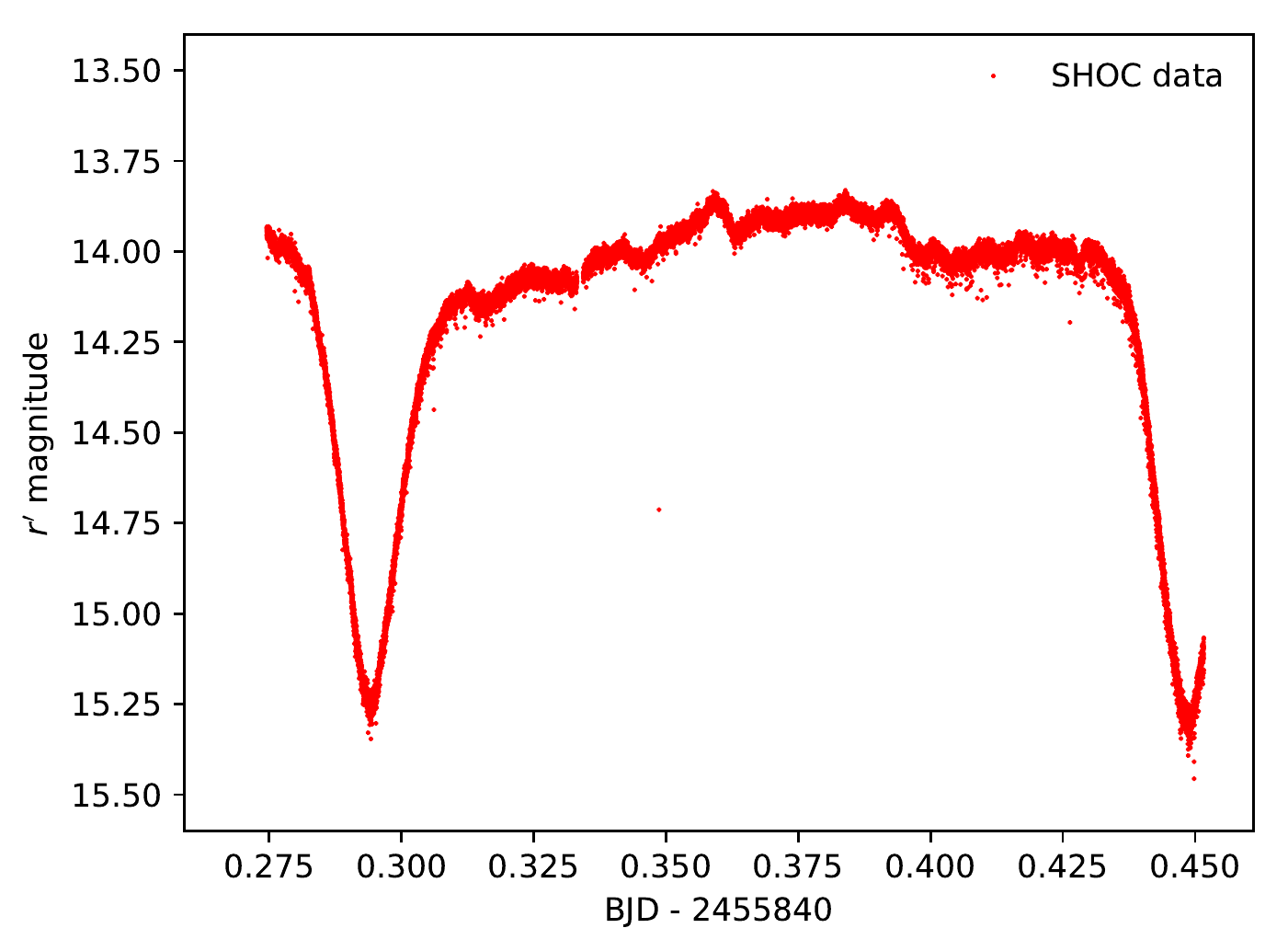}
\caption{Photometric light curve of EC21178-5417 obtained from our 2011 October 5 observations, showing the deep eclipse.}
\label{figure:light1}
\end{figure}

Follow-up photometric observations of EC21178-5417 were made with the SAAO 1.9-m telescope on the night of 2011 October 5 using the Sutherland High-speed Optical Camera (SHOC; \citealt{2013PASP..125..976C}) for the purpose of refining the ephemeris. A record of observation is given in Table \ref{tab:photo_obs}. 
EC21178-5417 was observed for a duration of 4.12 hours, and a total of two eclipses, one complete and the other a partial eclipse, were covered by the observations. 
The observations were made in white light (WL, unfiltered) mode. The SHOC camera was mounted below the filter wheel of the 1.9-m reflector telescope. 
The field of view is 1.29 $\times$ 1.29 arcmin$^2$ (see Fig. \ref{figure:finder}), and the data were binned 8 $\times$ 8 for a plate scale of 0.61 arcsec pixel$^{-1}$. Sky flat-fields were obtained at the beginning of the night with the same binning of the CCD. The exposure time was selected to be 0.5 s and the detector was used in frame transfer mode. The 1 MHz electron-multiplying (EM) mode was used with an EM gain of 20 and a pre-amplifier gain of 7.5 $e^{-}$pixel$^{-1}$. 
An effective read noise $\sim$1 $e^{-}$pixel$^{-1}$ is achieved by using these settings. The camera was thermo-electrically cooled at -70$\dg \,\rm{C}$, resulting in a dark current of $<$ 0.001 $e^{-}$pixel$^{-1}$. 

The photometric data of EC21178-5417 was reduced using the \texttt{IRAF}\footnote{IRAF is distributed by the National Optical Astronomy Observatories,  which  are operated by  the Association of Universities for Research in Astronomy, Inc., under cooperative agreement with the NSF.} data reduction package. EC21178-5417 and a comparison star were identified on each image with the \texttt{IRAF} task \texttt{DAOFIND}, and aperture photometry was obtained for both stars, using \texttt{PHOT} in the \texttt{DAOPHOT} package. 
Multiple apertures were used and an aperture correction was applied using \texttt{MKAPFILE} in \texttt{IRAF} \citep{1990PASP..102..932S}. 
We determined the brightness of the comparison star as $r'$ = 13.07 mag from simultaneous $g'$, $r'$, and $i'$ photometry using the three-colour imaging camera (TRIPOL) on the SAAO 0.75-m telescope. 

For the final light curve, Fig. \ref{figure:light1}, a differential correction was applied to the photometry of EC21178-5417 by assuming that the comparison star in the field is constant and has a brightness of $r'$ = 13.07 mag. Even though our observations are taken in white light mode, the details of variability and eclipse depth of EC21178-5417 match that of the lightcurve in $r'$ band obtained by Dr Nagayama with TRIPOL at the same time. Note that these observations were taken for the purpose of determining the precise time of mid-eclipse so precise photometry calibration was not essential. 
A total of two eclipses were obtained from the 2011 observations. 

At the end of the observation campaign, a total of 18 eclipses were obtained from these observations and we measured their mid-eclipse times and the results are listed in Table \ref{tab:med-eclipse}. All the mid-eclipse times were corrected for the light travel-time effect to the barycentre of the solar system (i.e. converted to the barycentric dynamical times (TDB) as barycentric julian dates (BJD; \citealt{2010PASP..122..935E})).

\subsubsection{Other observations}

EC21178-5417 was observed on eight separate nights between 2015 July 14 and 2016 May 12 (see \citealt{2017NewA...56...60B}, for details of these observations). A total of ten eclipses were observed (see Table \ref{tab:med-eclipse}), they measured ten mid-eclipse times. 
A further eclipse of EC21178-5417 was observed on 2016 July 19 with the SHOC instrument on the SAAO 1.9-m telescope (see \citealt{2020MNRAS.491..344R} for more details on this observations). Their mid-eclipse time is given in Heliocentric Julian Date and we converted this to BJD. We combined all the mid-eclipse times from the literature (e.g. \citealt{2017NewA...56...60B,2020MNRAS.491..344R})  with our 18 mid-eclipse times as to bring the final total to 29 mid-eclipse times and they are shown in Table \ref{tab:med-eclipse}.  

\subsection{Spectroscopy}

Observations of EC21178-5417 were made with the SAAO 1.9-m telescope using the Grating Spectrograph over nine nights stretching from 2002 September 7 until 2013 October 22. Table \ref{tab:1} provides our spectroscopic observing log. Our spectroscopic observations were made during good sky conditions, i.e. good seeing and low humidity.
The SITe CCD was used and the spectra were prebinned 1 $\times$ 2. A slit width of 1.5$''$ was used. Grating 6 was used for most of the observations (2002, 2011 October and 2013) with a central wavelength of $\sim$4400\AA{}, and a wavelength range of 3500--5400\AA{}. The grating angle was set to $\sim$14.1$\dg$, giving a reciprocal dispersion of 1.11 \AA{}/pixel. 
For the spectra obtained on 2011 September 5, a lower resolution grating (Grating 7) was used so that the central wavelength was $\sim$5700\AA{}, with a wavelength range of 3600--7600\AA{}. The grating angle was set to $\sim$17.2$\dg$, giving a reciprocal dispersion of 2.29 \AA{}/pixel. This includes the H$\alpha$ 
line, and telluric lines towards the red, as well as the spectral range previously covered. 
The exposure time for each spectrum was selected to be 600 s, or 900 s for the various observing runs 
(see Table \ref{tab:1}). These exposure times allow a sufficient phase resolution and coverage of the orbital cycle. 
For the purposes of flux calibration, the following standard stars were observed: EG21, LTT 7987 and 7379, and HR1544. 
Several spectra were taken in such a way that they cover one orbital cycle of the binary on each night, each separated 
by one or two Cu-Ar arc spectra with the exposure time of $\sim$30 s.

\begin{table*}
\caption{Spectroscopic observation log of EC21178-5417.}
\label{tab:1}                                                       
\medskip
\begin{center}
\begin{tabular}{l c c c c c c} \hline \hline
Date        & Number &Exposure& Grating  & Central     & Wavelength& Orbital \\
  of        &   of   & time   & angle    & wavelength  & range     & phase \\ 
observation & spectra&  (s)   & ($\dg$)  & (\AA{})     & (\AA{})   & coverage  \\ \hline \hline
2002 Sept 7* & 7      & 600   &  14.1    & 4400        & 3500-5400 & 0.76-1.25 \\ 
2011 Sept 5  & 24     & 600   &  17.2    & 5700        & 3550-7500 & 0.41-1.61 \\ 
2011 Oct 22  & 12     & 900   &  14.1    & 4400        & 3500-5400 & 0.85-1.72 \\ 
2011 Oct 23  & 12     & 900   &  14.1    & 4400        & 3500-5400 & 0.30-1.15 \\  
2013 Oct 16  & 20     & 900   &  14.1    & 4400        & 3500-5400 & 0.58-1.97 \\
2013 Oct 17  & 14     & 900   &  14.1    & 4400        & 3500-5400 & 0.94-1.87 \\ 
2013 Oct 18  & 20     & 900   &  14.1    & 4400        & 3500-5400 & 0.41-1.79 \\ 
2013 Oct 19  & 22     & 900   &  14.1    & 4400        & 3500-5400 & 0.85-2.35 \\   
2013 Oct 22  & 18     & 900   &  14.1    & 4400        & 3500-5400 & 0.34-1.56 \\  \hline
             & \\
\end{tabular}
\end{center}
Notes: The star * sign marks the observations retrieved from the archive.  
\end{table*}

\smallskip
The spectroscopic data of EC21178-5417 were reduced following standard procedures using 
\texttt{IRAF}\footnote{\texttt{IRAF} is distributed by the National Optical Astronomy Observatories, which are operated 
by the Association of Universities for Research in Astronomy, Inc., under cooperative agreement with the National 
Science Foundation} data reduction package. The frames were first overscanned, trimmed and then flat-fielded with a 
master flat-field. The spectra were extracted using \texttt{IRAF}'s implementation of optimal extraction using 
\texttt{APALL} task. The arc spectra were extracted at the position of the corresponding object spectrum and were then 
used to wavelength calibrate the target spectra by interpolation between neighbouring arcs. The spectra were corrected for 
instrumental response using the flux standard stars mentioned above. All the spectra were transformed to 
heliocentric rest frame and corrected for local atmospheric extinction. 
The final average one-dimensional spectrum for each date are presented in Fig.
\ref{figure:all}.

\section{Photometry}
\label{sec:photometry}

\subsection{Photometric ephemeris}\label{sec:ephe}

\begin{figure}
\centering
\includegraphics[width = 8.0cm]{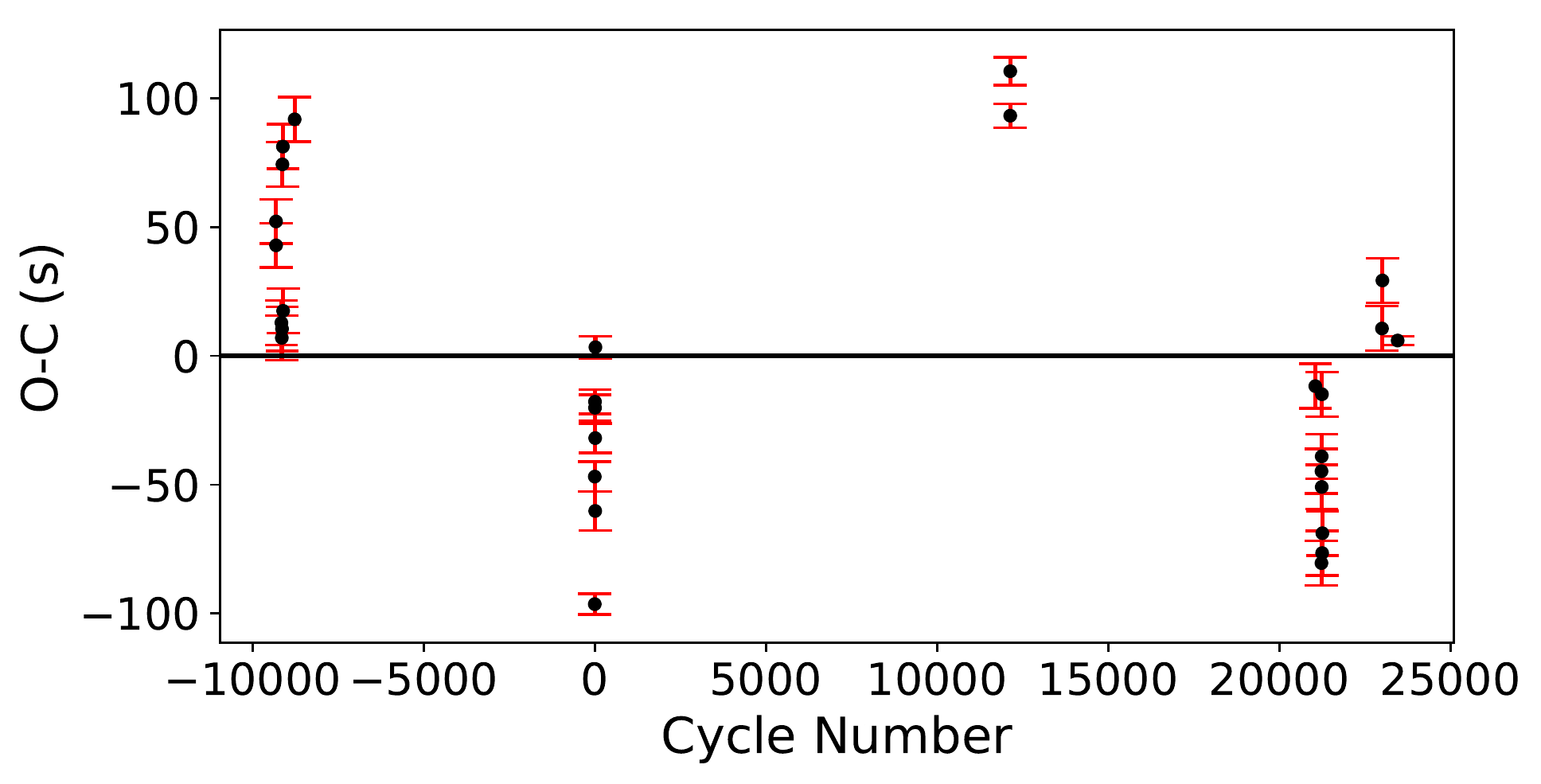}
\caption{ Observed minus calculated eclipse time residuals after subtraction of the eclipse ephemeris (Eq. \ref{eph1}). }
\label{figure:oc}
\end{figure}

The light curve of EC21178-5417 (Fig. \ref{figure:light1}) obtained on 2011 October 5 shows deep symmetrical V-shaped primary eclipse minima, with out-of-eclipse magnitudes of $\sim$13.8 mag and a depth of the primary eclipse of about 1.3 mag. 
The average magnitude out of eclipse of EC21178-5417 is consistent with previous observations made for this system in 2002 and 2006 \citep{2003MNRAS.344.1193W,2008.....Z}. 
The total width of the eclipse is $\sim$0.1 in orbital phase. There is some short time-scale variability present in the light curve, but this appears to be mere flickering. No DNOs or QPOs (reported by  \citealt{2003MNRAS.344.1193W}) were detected in the 2011 October observations.   

We used the mid-eclipse time from 2006 October 16 as the starting point, $T_0$, and used the orbital period from \cite{2008.....Z} to calculate the cycle numbers. Table \ref{tab:med-eclipse} gives all the 29 mid-eclipse times of EC21178-5417. After using a linear least squares fit to the above mentioned timings, we determined the following orbital ephemeris: 

\begin{equation}
\label{eph1}
\begin{split}
 \rm{BJD_{min}} = 2453964.3323\,\ (\pm 1 \times 10^{-4})\,\ + \\ 0.15452724\,\ (\pm 1 \times 10^{-8})\rm{E}.
\end{split}
\end{equation}

\noindent
Figure \ref{figure:oc} shows the  observed minus calculated (O--C) diagram after subtraction of the best fit (straight line) ephemeris based on the data outlined in Table \ref{tab:med-eclipse}. 
The best fitting results to the residuals can be obtained by using a quadratic or cubic ephemeris to the mid-eclipse times. This will significantly reduce the residuals seen in Fig. \ref{figure:oc}. 
We note that the scatter of eclipse times within each epoch and the total time span of our data set does not warrant additional terms. 
The individual error bars in Fig. \ref{figure:oc} are formal errors from fitting a Gaussian to the eclipse profile. 
The scatter within each epoch suggests that variability (e.g. flickering) is adding an additional random contaminant to measuring the true centre of the eclipse. 
This ephemeris (Eq. \ref{eph1}) was used to phase all of our spectroscopic data.

\begin{table}
\caption{Eclipse timings of EC21178-5417 (given in BJD-TDB - 2450000) and cycle numbers. }
\label{tab:med-eclipse}
\begin{center}
\begin{tabular}{llccc} \hline \hline
 Date of     &  Time of     & Error in&  Cycle  & Reference \\
 start       &  mid-eclipse & time of &  Number &  \\
 of night    &  (2400000+)  & eclipse &         &  \\  \hline \hline
2002 Sept 7  & 52525.375126 & 0.0001 & -9312 & $^a$  \\  
2002 Sept 7  & 52525.529546 & 0.0001 & -9311 & $^a$  \\  
2002 Oct 1   & 52549.480922 & 0.0001 & -9156 & $^a$  \\  
2002 Oct 3   & 52551.335181 & 0.0001 & -9144 & $^a$  \\  
2002 Oct 4   & 52552.416912 & 0.0001 & -9137 & $^a$  \\  
2002 Oct 6   & 52554.271978 & 0.0001 & -9125 & $^a$  \\  
2002 Oct 8   & 52556.280912 & 0.0001 & -9112 & $^a$  \\  
2002 Oct 9   & 52557.361865 & 0.0001 & -9105 & $^a$  \\  
2002 Nov 30  & 52609.283881 & 0.0001 & -8769 & $^a$  \\  
2006 Aug 16  & 53964.33115  & 0.000047 & 0      & this work \\
2006 Aug 16  & 53964.48625  & 0.000067 & 1      & this work \\
2006 Aug 17  & 53965.41375  & 0.000055 & 7      & this work \\
2006 Aug 17  & 53965.56825  & 0.000058 & 8      & this work \\
2006 Aug 18  & 53966.34075  & 0.000066 & 13     & this work \\
2006 Aug 18  & 53966.49495  & 0.000087 & 14     & this work \\
2006 Aug 19  & 53967.422849 & 0.000049 & 20     & this work \\
2011 Oct 5   & 55840.294147 & 0.000054 & 12140  & this work \\
2011 Oct 5   & 55840.448878 & 0.000062 & 12141 & this work \\
2015 Jul 14  & 57217.74883  & 0.0001 & 21054 & $^b$  \\
2015 Aug 11  & 57245.562940 & 0.0001 & 21234 & $^b$  \\
2015 Aug 11  & 57245.717880 & 0.0001 & 21235 & $^b$  \\
2015 Aug 12  & 57246.64511  & 0.0001 & 21241 & $^b$  \\
2015 Aug 12  & 57246.79950  & 0.0001 & 21242 & $^b$  \\
2015 Aug 13  & 57247.72708  & 0.0001 & 21248 & $^b$  \\
2015 Aug 14  & 57248.65353  & 0.0001 & 21254 & $^b$  \\
2015 Aug 15  & 57249.73531  & 0.0001 & 21261 & $^b$  \\
2016 May 10  & 57518.76817  & 0.0001 & 23002 & $^b$  \\
2016 May 12  & 57520.77724  & 0.0001 & 23015 & $^b$  \\
2016 Jul 19  & 57589.541596 & 0.00002  & 23460 & $^c$ \\
\hline \hline
\end{tabular}
\end{center}
Notes: $^a$\cite{2003MNRAS.344.1193W}, $^b$\cite{2017NewA...56...60B} and $^c$\cite{2020MNRAS.491..344R}. 
\end{table}

\subsection{System's inclination and dimensions}
\label{sec:geometry}

We now use the eclipse length and the orbital period of the system in order to estimate the mass ratio, inclination as well as the masses of the stellar components. 
In order to estimate the orbital inclination, we consider the geometry of a point eclipsed by a spherical body. It is possible to determine the inclination, $i$, of a binary system through the relation 

\begin{equation}
\left(\frac{R_2}{a}\right)^2 = \rm{sin}^2(\pi \Delta\varphi_{1/2}) + \rm{cos}^2(\pi \Delta\varphi_{1/2}) \rm{cos}^2 \textit{i}, 
\end{equation}

\noindent
where $R/a$ is the volume radius of the secondary star and $\Delta\varphi_{1/2}$ is the mean phase full-width of the eclipse at half the out-of-eclipse intensity. According to Eq. (2) of \cite{1983ApJ...268..368E}, the volume radius of the secondary star depends only on the mass ratio, $q$ = $M_2/M_1$, as shown below: 

\begin{equation}
    \frac{R_2}{a} = \frac{0.49\,q^{2/3}}{0.6\,q^{2/3} + \rm{ln}(1 + \textit{q}^{1/3})}. 
\end{equation}

\noindent
We calculated $\Delta\varphi_{1/2}$ from the individual light curves covering complete eclipse. We took the average of the eclipse widths and the derived value is $\Delta\varphi_{1/2}$ = 0.092 $\pm$ 0.003. 

The mass of the secondary star can be estimated using the mean empirical mass-period relationship, that is Eq. 9 from \citet{1998MNRAS.301..767S} as follows: 
\begin{equation*}
    M_2(M_{\odot}) = (0.126 \pm 0.011)P - (0.11 \pm 0.04) = 0.36(\pm 0.08) 
\end{equation*}
where $P$ is the orbital period expressed in hours. 
For stable mass transfer we require that $M_2/M_1$ < 1 and, therefore, using the above estimate of $M_2$ (= 0.36 $M_{\odot}$), the mass of the primary must be within the range 0.36 < $M_1$ < 1.44 $M_{\odot}$. This as an equivalent of 0.25 < $q = M_2/M_1$ < 1. 
However, according to \cite{1997PhDT........28G} and \cite{1998MNRAS.301..767S}, the average mass of the WD in CVs above the period gap is $\sim$0.8 $M_{\odot}$ (e.g. HS 0728+6738 \citealt{2004A&A...424..647R}). 
In order for us to provide an estimate of the uncertainty for the orbital inclination, we will assume that the true mass of the primary lies in the range 0.6-1.0 $M_{\odot}$ (i.e. 0.36 < $q$ < 1.0). 
Taking this into account we obtain an orbital inclination of 83 $\pm$ 7$^{\circ}$. The uncertainty in $i$ include systematic ones from the use of mass period relation and from the assumption of an axially symmetric accretion disc. The statistical uncertainties in the measurement of $\Delta\varphi_{1/2}$ is then negligible.

\section{Spectral Analysis}
\label{sec:SpectraAnalysis}

\subsection{Averaged spectra of EC21178-5417}
\label{sec:spectra1}

\begin{figure*}
\centering
\includegraphics[width = 0.95\textwidth ]{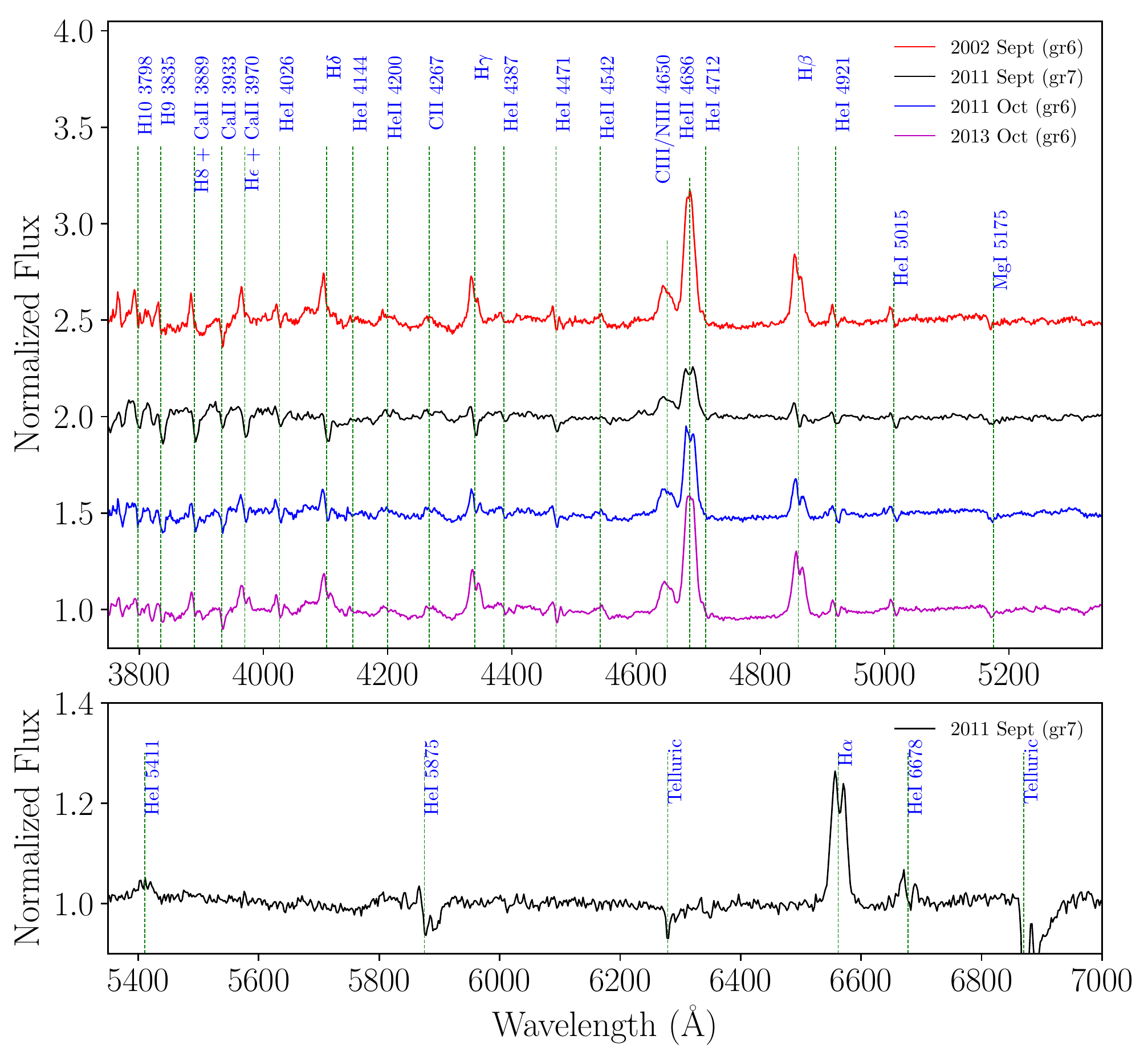}
\caption{Averaged continuum-normalized spectra of EC21178-5417, in which each individual spectrum is given equal weighting, obtained from four different dates. Top panel: (from top to bottom) spectrum from  2002 September (red), 2011 September (black), 2011 October (blue) and 2013 October (magenta). Each spectrum in the top panel is shifted up by increment of 0.5 from the bottom to top. The bottom panel shows the red averaged spectrum from 2011 September. The prominent lines have been identified and labelled in each panel.}
\label{figure:all}
\end{figure*}

The averaged continuum-normalized spectrum of EC21178-5417 obtained over four different epochs is shown in Fig. \ref{figure:all} (top and bottom panels), and is quite remarkable in its detail and wealth of features. 
The blue spectrum (Fig. \ref{figure:all} top panel) is dominated by the emission line from HeII 4686 \AA{} (strongest feature). The strength of the HeII 4686 \AA{} emission line is typical of a NL system with a high rate of mass-transfer \citep{2004AJ....128.1882S}. 
The Balmer lines (H$\beta$, H$\gamma$, H$\delta$ down to H$10$) are clearly present with an increasing absorption component towards the blue end of the spectrum. Balmer emission was more pronounced in 2011 October in H$\beta$, H$\gamma$, H$\delta$, and H$\epsilon$. 
The HeI lines (at 4026 \AA{}, 4471 \AA{}, 4921 \AA{}, and 5015 \AA{}) all have a strong absorption component. The Bowen (CIII/NIII) blend at 4640--4650 \AA{} and CII at 4267 \AA{} are clearly seen in emission. 
In addition, the blue spectrum shows absorption and emission of HeII (at 4200 \AA{} and 4542 \AA{}, part of the HeII Pickering series) as well as absorption from MgI at 5175 \AA{}. Also visible are the CaII lines (at 3933 \AA{} and 3969 \AA{}) towards the blue part of the spectrum in absorption. 

The red spectrum (Fig. \ref{figure:all}, bottom panel) is dominated by double-peaked H$\alpha$ emission: a typical signature of disc emission, due to the Doppler shifting of orbiting disc material; HeII 5411 \AA{} and, HeI at 5875 \AA{} and 6678 \AA{} are seen in emission with an absorption core. 
There is a clear evidence for an asymmetry in the line profile (more specifically in H$\alpha$ line), suggesting that emission from another location within the system contributes. 
This has been seen in other CVs (e.g. detached WD/M-dwarf binaries) in low states, where they originate either on or close to the WD \citep[e.g.][]{2007A&A...474..205T,2011A&A...532A.129T,2012MNRAS.419..304P,2013MNRAS.436..241P} or from material located between the two stars \citep[e.g.][]{1998A&A...338..933G,2003MNRAS.345..506O,2011MNRAS.412.2563P}.

\medskip 
In the top panel of Fig. \ref{figure:all} we show and compare the average continuum-normalized blue spectra of EC21178-5417 obtained from the four different dates: 2002 September (red), 2011 September (black), 2011 October (blue) and 2013 October (magenta). 
The average spectrum of EC21178-5417 from 2002 September (Fig. \ref{figure:all}) is dominated by emission from HeII 4686 \AA{} and the Balmer lines with very little absorption from HeI lines. This is similar to the average spectrum obtained from the 2013 October observations (Fig. \ref{figure:all}). 
This suggests that EC21178-5417 was in a similar state during the 2002 September and 2013 October observations. 
On the other hand, the average spectrum obtained in 2011 September (Fig. \ref{figure:all}) show strong absorption from the Balmer (from H$\gamma$ down to H$10$) and HeI lines (e.g., HeI 4471 \AA{}). Even the H$\beta$ itself has a strong absorption component. This suggest that EC21178-5417 was caught in a distinct state in 2011 September compared to that of 2002 September and 2013 October. 
The average spectrum obtained in 2011 October (Fig. \ref{figure:all}) exhibits both absorption and emission features. Noticeable are broad and double-peaked emission features from H$\beta$ and H$\gamma$, which suggest that these lines originate in an accretion disc. The blueshifted peaks of H$\beta$ and H$\gamma$ are stronger than the redshifted peaks. 
This effect has been observed in other CVs such as the dwarf nova V2051 Oph \citep{2001MNRAS.323..484S} but they did not attempt an interpretation other than that the blue side of the accretion disc could be making a larger contribution to the emission. The H$\delta$ and H$\epsilon$ lines show similar behaviour to that of H$\beta$ and H$\gamma$, but for the former, the redshifted peak appears in absorption. 

The averaged spectra of EC21178-5417 show distinct spectral features depending on various factors such as the state of the accretion disc, mass transfer rate, etc, which seems to vary on timescales of months and years (see Fig. \ref{figure:all}). For example, the 2011 October spectra (Fig. \ref{figure:all}) shows the intermediate state, where both emission and absorption are present. Thus this system changes from mostly emission (2002 September), to absorption (2011 September), and then to intermediate state a month later in 2011 October. But two years later, 2013 October, the average spectrum is again dominated by emission. 
This is somewhat unexpected given that the out-of-eclipse brightness of EC21178-5417 is similar throughout the individual observation runs. This could be attributed to a change in mass transfer rate between the individual observations but could also be due to the varying strength of the absorption component. The other spectral features that are present also vary from one observation to the other.


\subsection{The bright spot spectrum}\label{sec:bs}

\begin{figure*}
\centering
\includegraphics[width = 0.95\textwidth ]{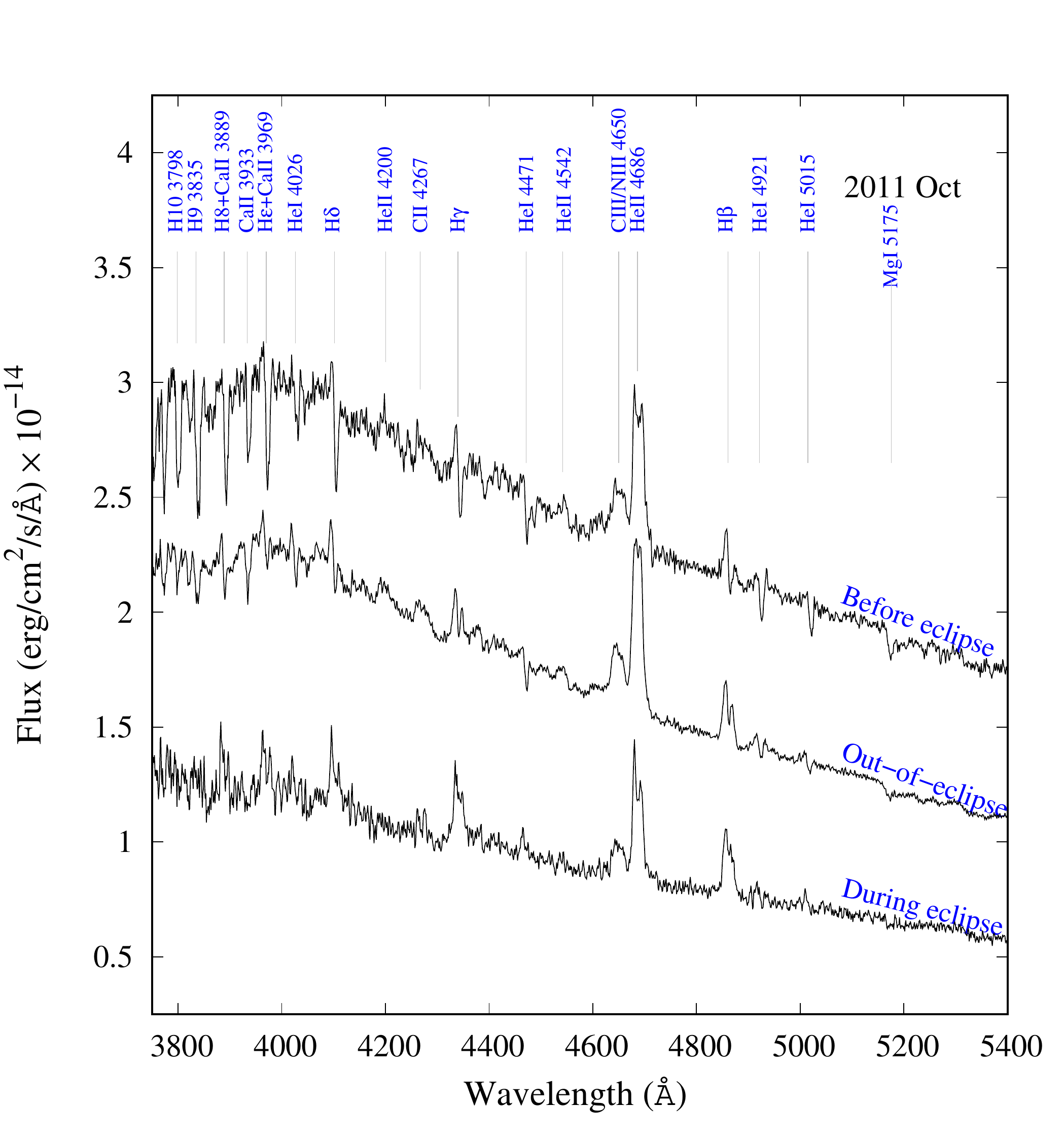}
\caption{Averaged spectra of EC21178-5417 at phase intervals 0.75 $\leq \phi <$ 0.95 (top, offset by +7.0 $\times 10^{-15}$), 0.05 $ \leq \phi <$ 0.75 (middle) and 0.95 $\leq \phi <$ 1.05 (bottom). The prominent lines have been marked and labelled. }
\label{figure:phase1}
\end{figure*}

Figure \ref{figure:phase1} shows the average flux-calibrated spectrum of EC21178-5417 at three different phase bins: before the eclipse (0.75 $\leq \phi <$ 0.95), during the eclipse (0.95 $\leq \phi <$ 0.05), and out-of-eclipse (0.05 $\leq \phi <$ 0.75). It is clear from the figure that the spectrum changes substantially with orbital phase. 
The continuum shape of the spectrum obtained during mid-eclipse (bottom of Fig. \ref{figure:phase1}) is slightly flatter than the other two; this was also noted by \citet{2008.....Z}. The Balmer series, HeII 4686 \AA{}, HeI lines and CIII/NIII blend are present and double-peaked in emission. 
The strength of the HeII 4686 line and the CIII/NIII blend at 4650\AA{} are reduced during eclipse, whereas that of the Balmer lines are enhanced. This indicates that the former lines originate close to the WD and the inner accretion disc, and they are more affected by the eclipse.

\smallskip
The out-of-eclipse spectrum of EC21178-5417 (Fig. \ref{figure:phase1}, middle) is similar to the average spectrum shown in Fig. \ref{figure:all} (top panel, blue line) with strong and broad double-peaked emission lines from HeII 4686 \AA{} and the Balmer lines. HeI lines and CIII/NIII blend are also present in broad emission and/or absorption on a steep continuum. The ratio of HeII 4686 \AA{} to H$\beta$ is larger than unity, and the presence of HeII 5411\AA{} suggests that these emission lines are formed in a region with a higher than usual level of ionization \citep{2001A&A...368..183G}. 

\smallskip
The spectrum of EC21178-5417 obtained before the eclipse (Fig. \ref{figure:phase1}, top) revealed that the higher Balmer lines (H$\epsilon$ to H$10$) change from general emission between phases 0.05 $\leq \phi <$ 0.75 to absorption at phases 0.75 $\leq \phi <$ 0.95. This has been observed during the low states of SW Sex \citep{2001A&A...368..183G} and UX UMa \citep{2011MNRAS.410..963N}, and is attributed to contribution from the bright spot. 
Thus the higher Balmer lines likely originate from the vicinity of the bright spot which is clearly visible at phases 0.75 $\leq \phi <$ 0.95 (see figure 15 of \citealt{2001A&A...368..183G}). The high excitation features, HeII 4686 \AA{}, CIII/NIII blend at 4650 \AA{} and CII 4267\AA{}, do not change in strength compared to the continuum and each other. 
However, H$\beta$, H$\gamma$ and H$\delta$, each have two components: the redshifted absorption peak and the blueshifted emission peak. But the absorption is stronger than the emission, which is the opposite of what was seen for H$\delta$ and H$\epsilon$ at earlier phases. Furthermore, the HeI lines also change from emission and broad absorption to strong absorption.

\section{Doppler tomography}
\label{sec:Doppler}

\subsection{Trailed spectra of the line profiles}\label{sec:aa}

The middle and right panels of Figs \ref{figure:Dop1} to \ref{figure:Dop2a} show continuum-normalized and phase-binned trailed spectra (observed and reconstructed) of HeII 4686 \AA{}, the Balmer lines and HeI 4471 \AA{}, respectively. The reconstructed spectra are constructed based on features shown in the Doppler maps. 
The agreement between the observed and reconstructed trailed spectra serves as a measure of the reliability of the Doppler maps. 
The trailed spectra of HeII 4686 \AA{} shows two peaks moving in anti-phase with respect to each other, whereas that of the Balmer lines (and HeI) shows two peaks moving in-phase with each other. 
 
\smallskip
The observed trailed spectra of HeII 4686 \AA{} lines from the 2011 observations (Figs \ref{figure:dopHeII1011} and \ref{figure:dopHeII0911}, middle panels) show two peaks in emission separated by $\sim$1000 km/s at phase 0.0 with the semi-amplitude of $\sim$600 km/s. The two peaks move in anti-phase direction with respect to each other. At phase $\sim$0.2, the two peaks come to within 200 km/s of each other; it is not clear if these two peaks swap side. 
But at about phase $\sim$0.7, the redshifted peak crosses the zero velocity from red to blue, whereas the blueshifted peak crosses the zero velocity from blue to red (see Figs \ref{figure:dopHeII1011} and \ref{figure:dopHeII0911} middle panels). 
Similarly, the middle panel of Fig. \ref{figure:dopHeII1013} shows the trailed spectra of HeII 4686 \AA{} obtained after subtracting the average spectrum from the individual spectra taken on 2013 October. This is the main reason why the trailed spectra of HeII 4686 \AA{} in Figs \ref{figure:Dop1} and \ref{figure:Dop2} appears different. 
The reconstructed trailed spectra for HeII 4686 \AA{} lines (right panels of Figs \ref{figure:dopHeII1011} and \ref{figure:dopHeII0911} show consistent results with the observed trailed spectra and are similar to those of IP Peg for the same line \citep{1999MNRAS.306..348H}. 
The only difference is the absence of the third central peak in emission (associated with the secondary star\footnote{This central feature is associated with the secondary star and is seen as an S-wave in the trailed spectra. 
It is most common in DNe during outbursts and is thought to be due to the irradiation of the secondary star, e.g. IP Peg \citep{1999MNRAS.306..348H}.}) in the trailed spectra of HeII 4686 \AA{} of EC21178-5417, which is present for IP Peg. The reconstructed trailed in the right panel of Fig. \ref{figure:dopHeII1013} is also consistent with the observed trailed spectra. 

\smallskip
Figure \ref{figure:dopHalpha0911} (middle panel) shows the observed trailed spectra of the H$\alpha$ line from 2011 September 5 data. There are two emission peaks that are clearly visible throughout the orbital cycle. For this line, the two emission peaks are moving in-phase with respect to each other. 
As with the HeII 4686 lines, the observed trailed spectra of the H$\alpha$ line do not show any emission that is attributed to the secondary star. This is evident by the absence of the central emission feature seen in the trailed spectra of IP Peg during outbursts \citep{1997MNRAS.290L..28S}. The reconstructed trailed spectra of this line is also consistent with the observed trailed spectra. 

\smallskip
The observed trailed spectra of the H$\beta$ line from 2011 October (Fig. \ref{figure:dopHbeta1011} middle panel) and 2013 October (Fig. \ref{figure:dopHbeta1013} middle panel) resembles that of the H$\alpha$ line. There are two emission peaks that are clearly visible and moving in-phase with each other. 
The central absorption dip is present in H$\beta$. However, at $\phi \sim$ 0.7, the second peak disappear or switches to absorption and re-appears at $\phi \sim$ 0.9. 
The reconstructed spectra of the H$\beta$ line (Figs \ref{figure:dopHbeta1011} and \ref{figure:dopHbeta1013} show two peaks moving in anti-phase with respect to each other. 
It is not clear if the two peaks crosses the zero velocity.

Lastly, the observed and reconstructed trailed spectra of H$\gamma$ are presented in Fig. \ref{figure:dopHgamma1013}  (middle and right panels), and shows similar results with those of the H$\beta$ line. 
The observed trailed spectra of HeI 4471 \AA{} is also shown in Fig. \ref{figure:dopHeI44711013}. It shows a mixture of absorption and emission and is difficult to interpret. The reconstructed spectra of HeI 4471 \AA{} does not reproduce the input spectra.

\subsection{Doppler maps}
\label{sec:dop}

\begin{figure*}
\centering
 \subfigure[HeII 4686 \AA{} line (2011 October 22--23)]{
   \includegraphics[width = 7.50cm ]{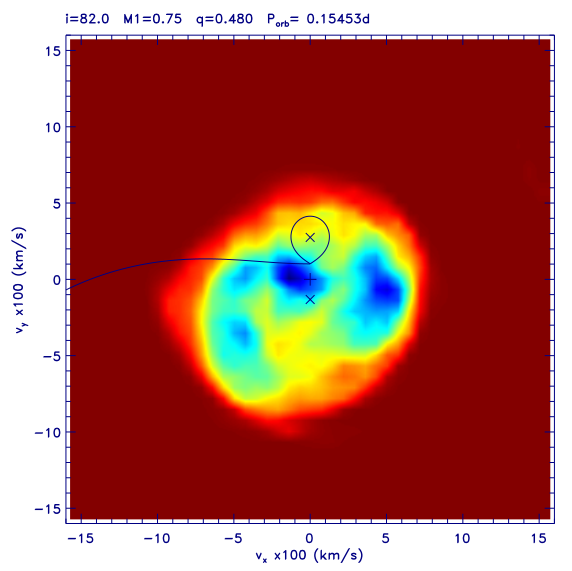}
   \hspace{0.5cm}
   \includegraphics[width = 4.0cm, height = 7.50cm]{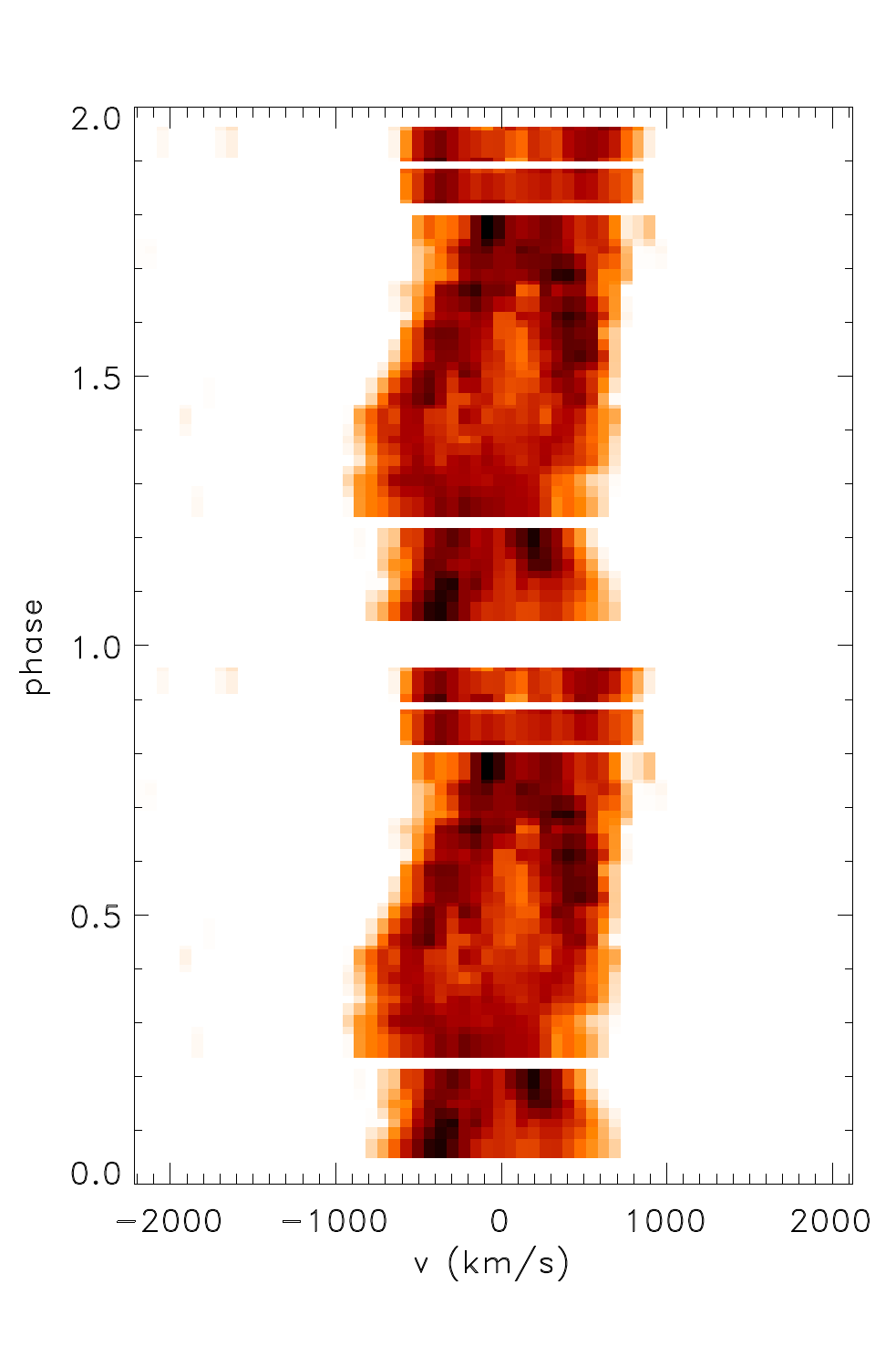}
   \includegraphics[width = 4.0cm, height = 7.50cm]{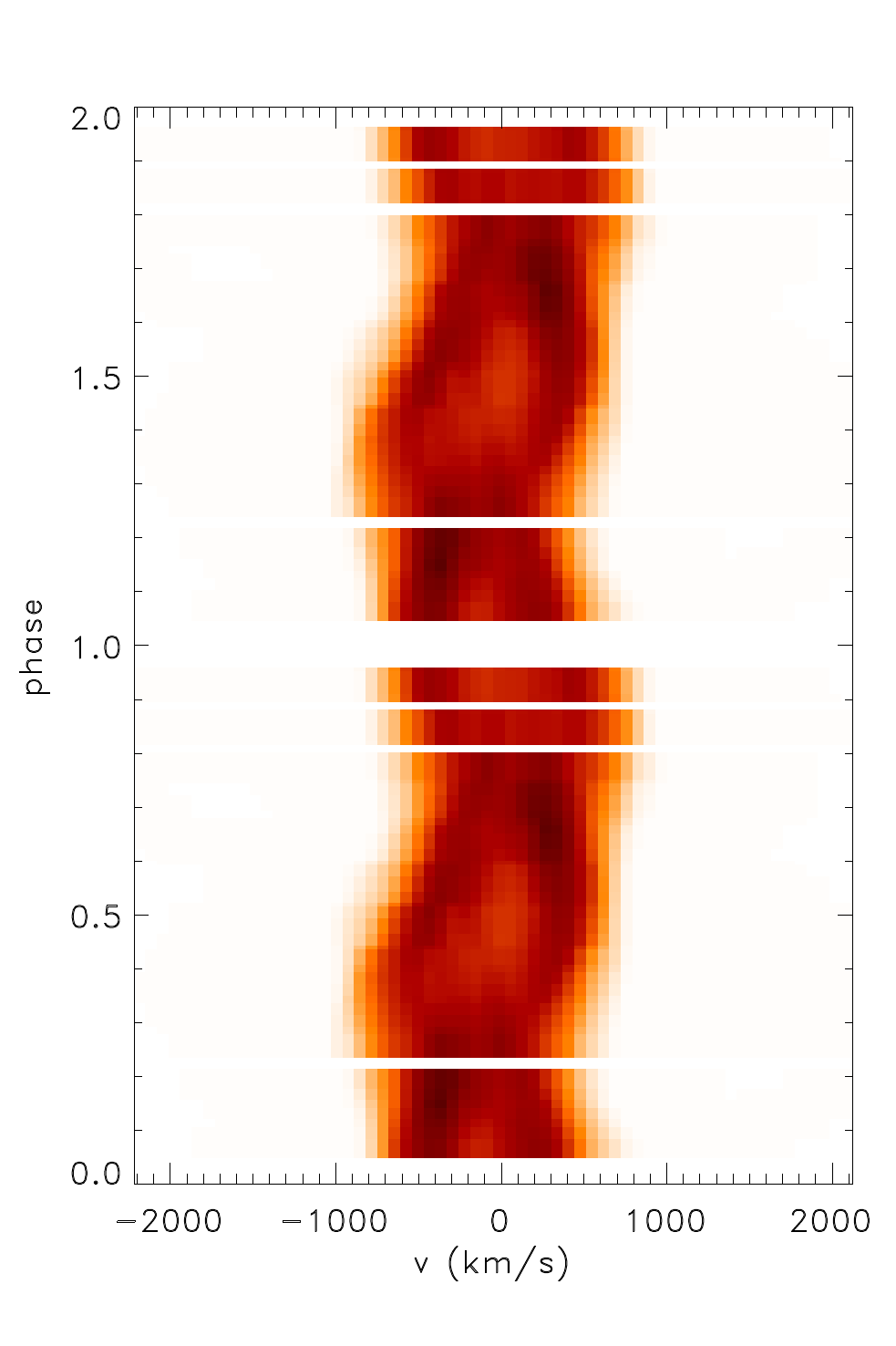}
   \label{figure:dopHeII1011}
   }
 \subfigure[HeII 4686 \AA{} line (2011 September 5)]{
   \includegraphics[width = 7.50cm ]{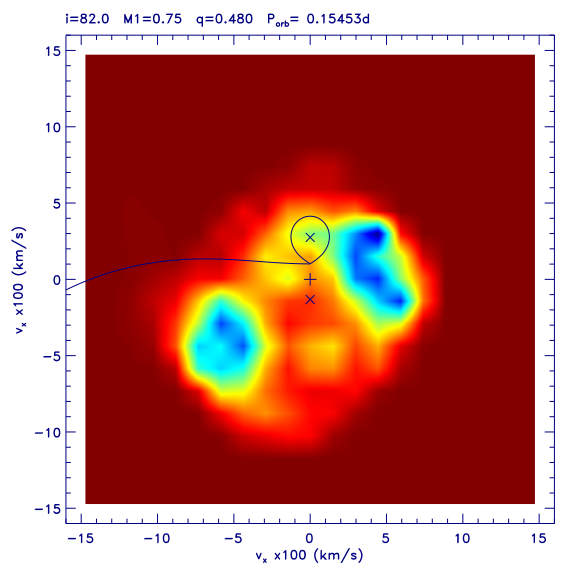}
   \hspace{0.5cm}
   \includegraphics[width = 4.0cm, height = 7.50cm]{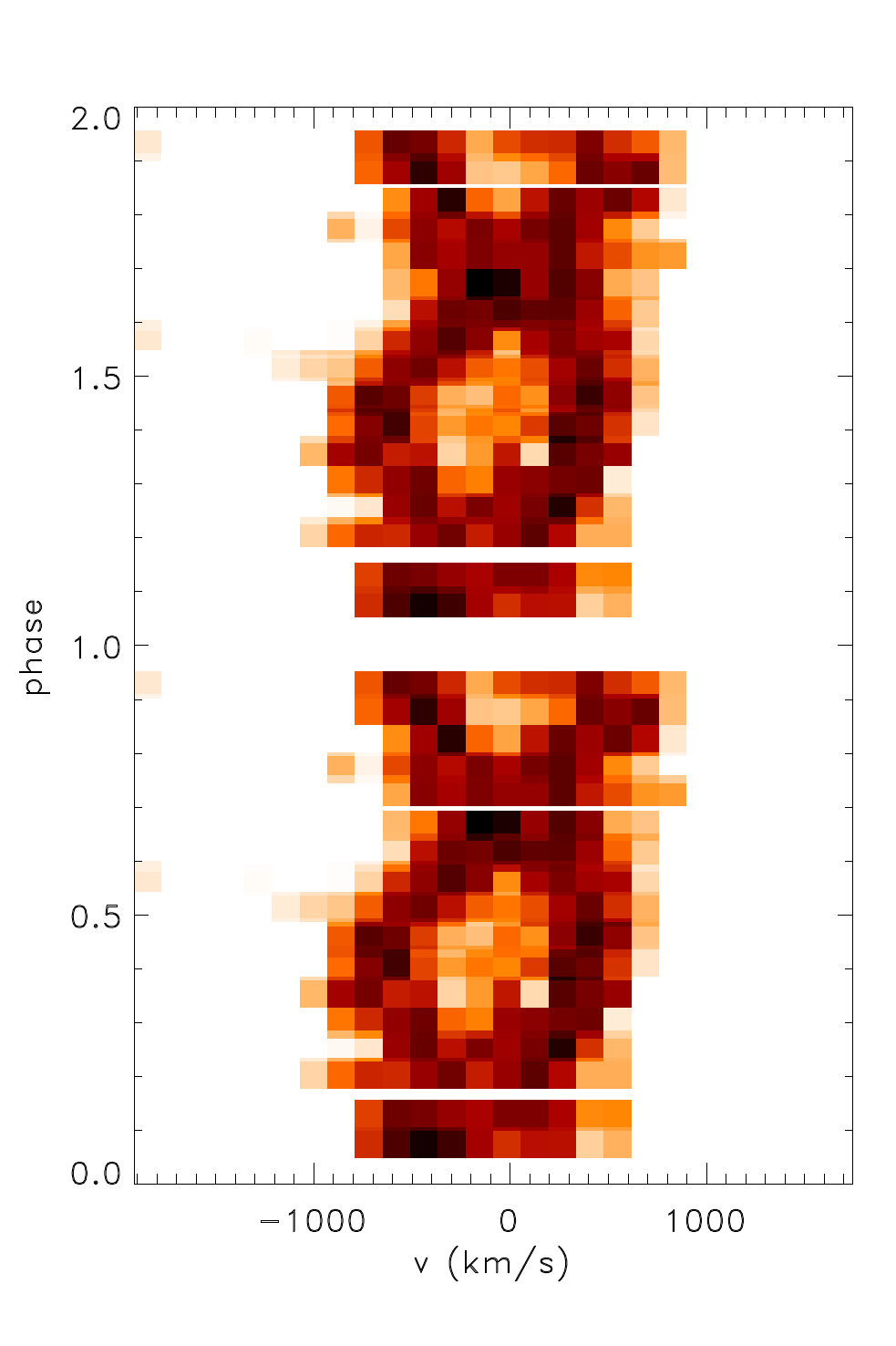}
   \includegraphics[width = 4.0cm, height = 7.50cm]{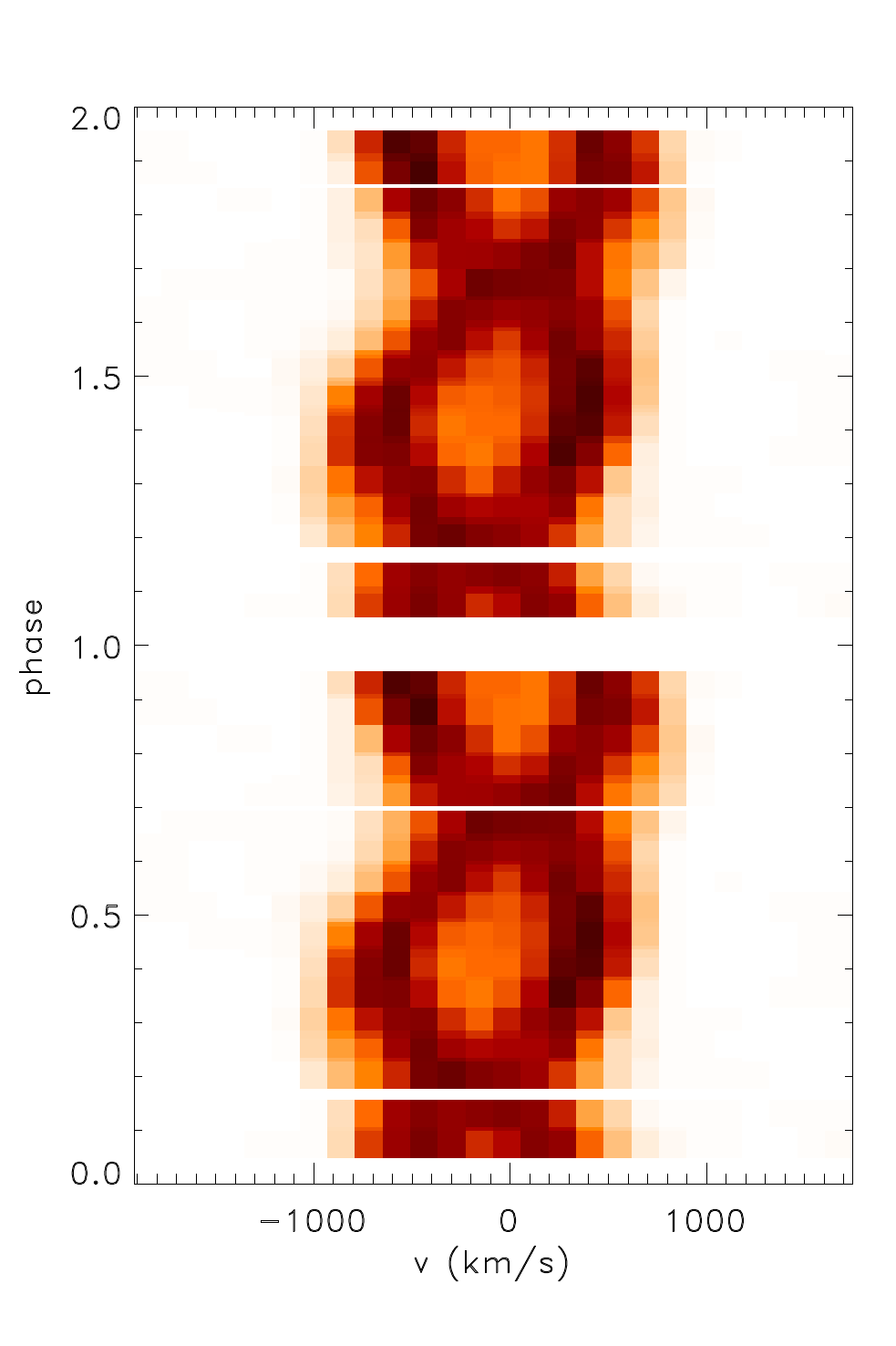}
   \label{figure:dopHeII0911}
   }
   \caption{Doppler maps (left panels) and trailed observed (centre panels) and reconstructed (right panels) spectra of HeII 4686 \AA{} emission lines.}
\label{figure:Dop1}
\end{figure*}

\begin{figure*}
\centering
 \subfigure[H$\alpha$ line (2011 September 5)]{
   \includegraphics[width = 7.50cm ]{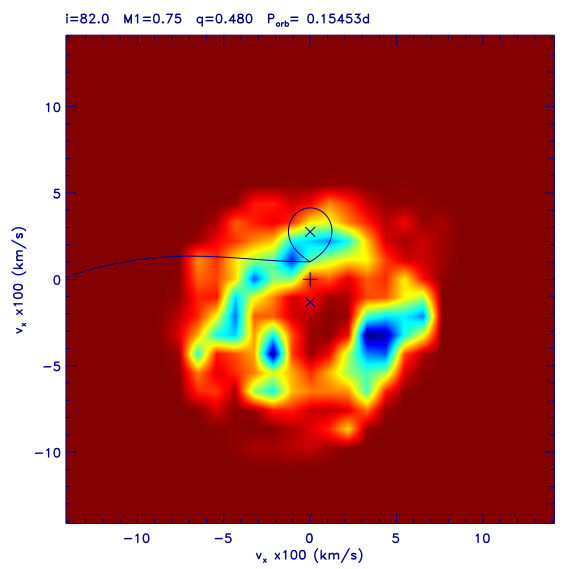}
   \hspace{0.5cm}
   \includegraphics[width = 4.0cm, height = 7.50cm]{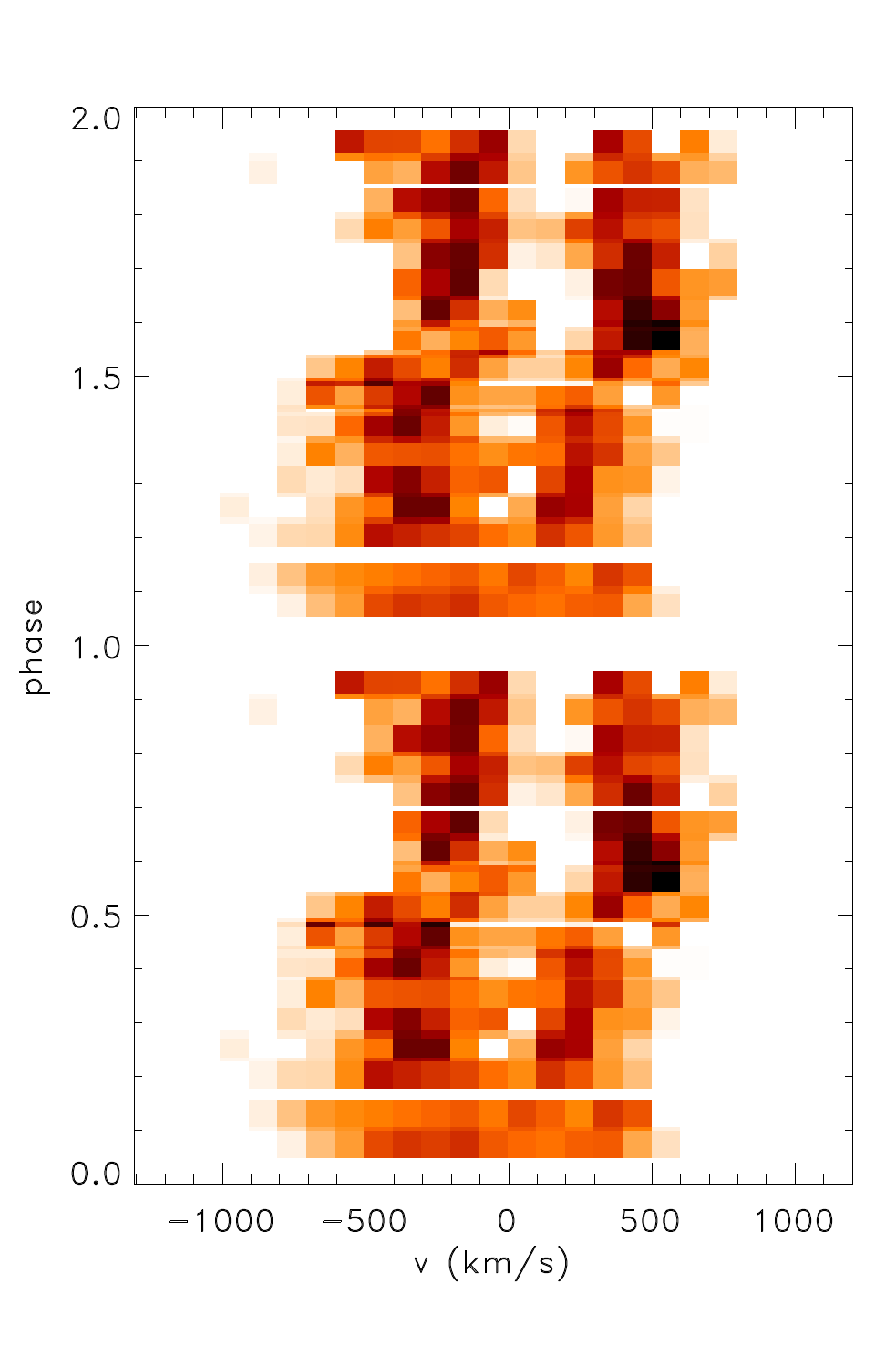}
   \includegraphics[width = 4.0cm, height = 7.50cm]{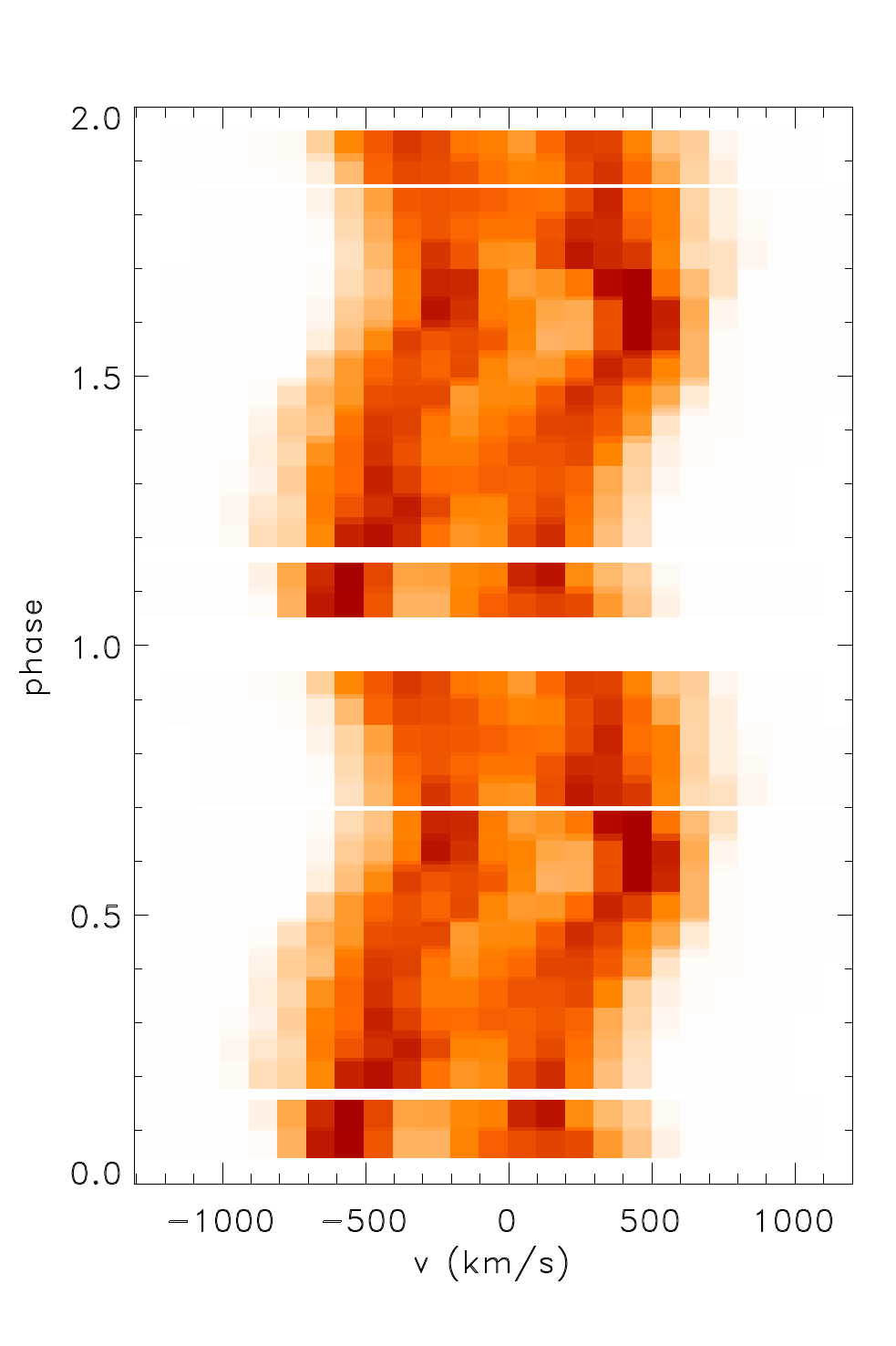}
   \label{figure:dopHalpha0911}
   }
 \subfigure[H$\beta$ line (2011 October 22--23)]{
   \includegraphics[width = 7.50cm ]{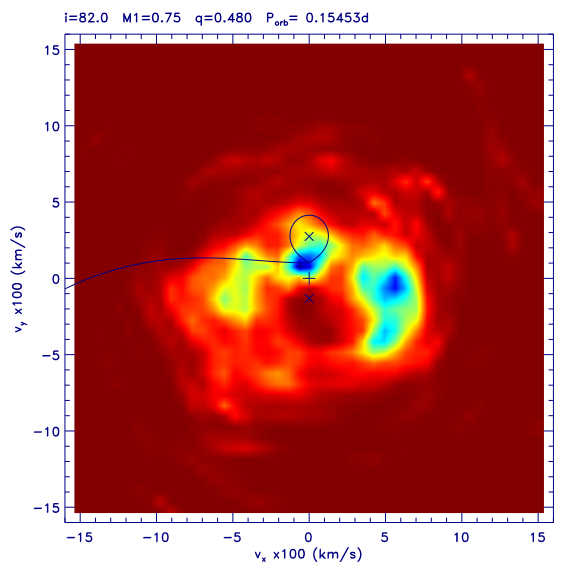}
   \hspace{0.5cm}
   \includegraphics[width = 4.0cm, height = 7.50cm]{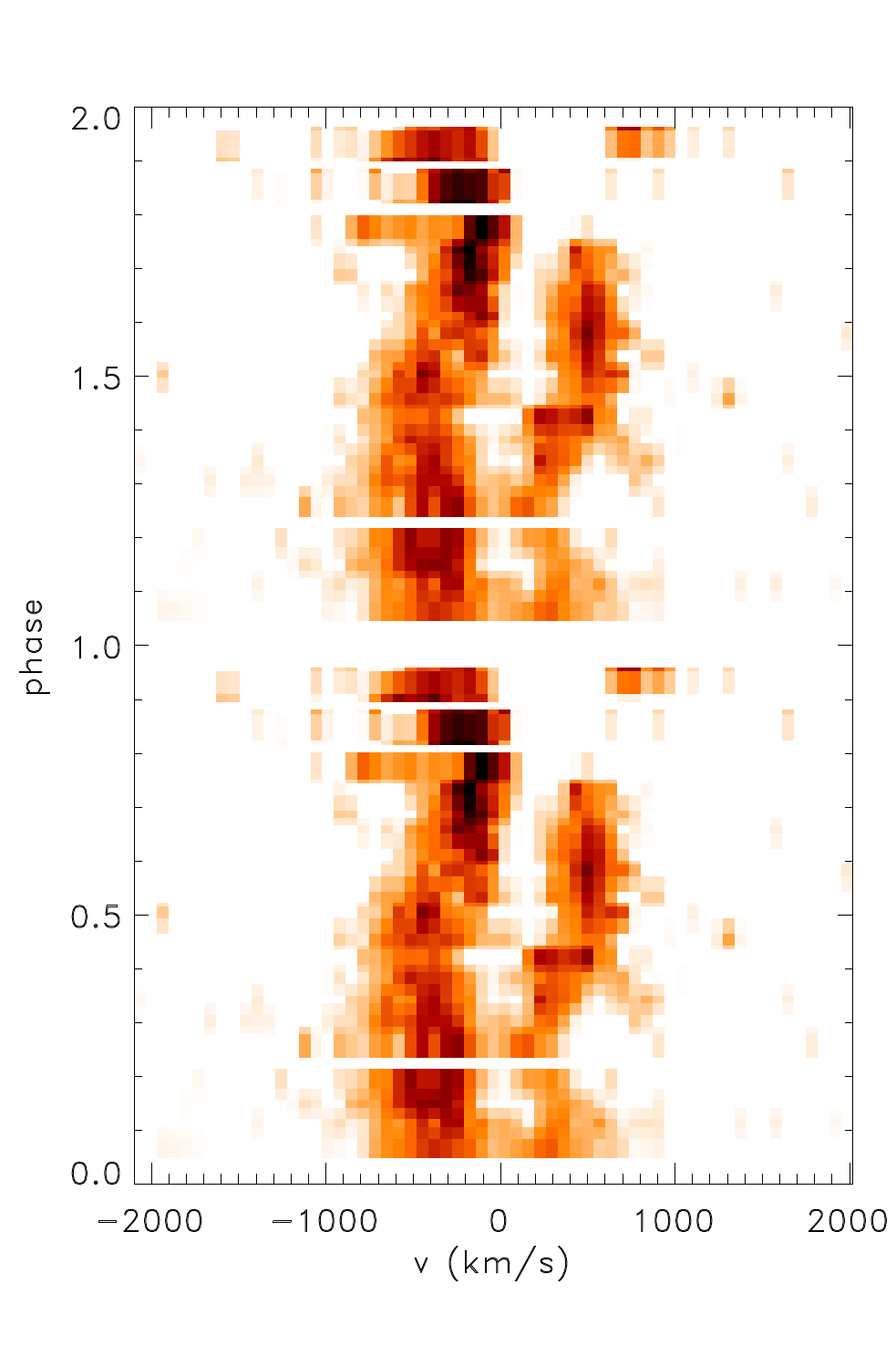}
   \includegraphics[width = 4.0cm, height = 7.50cm]{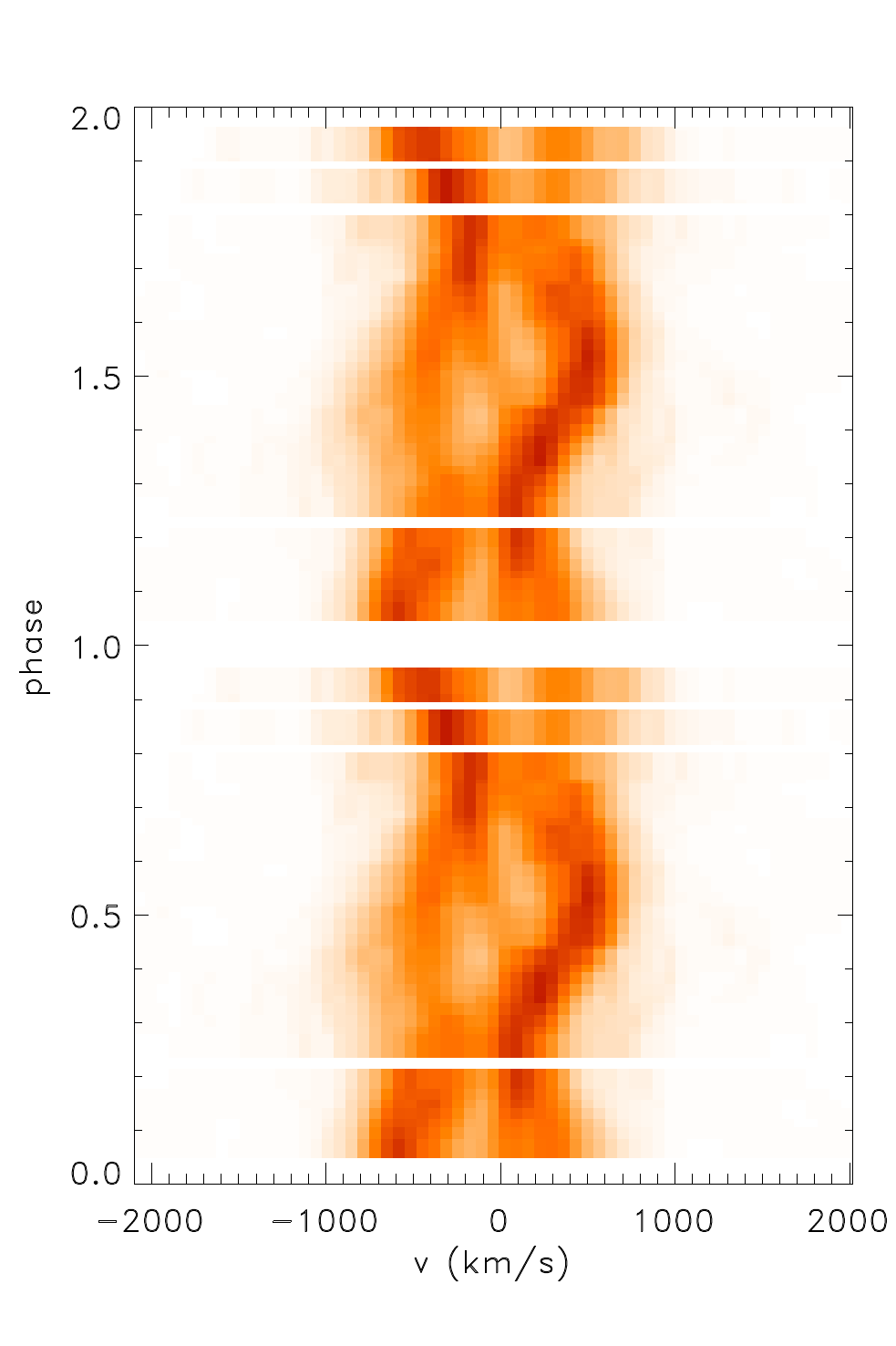}
   \label{figure:dopHbeta1011}
   }
\caption{Doppler maps (left panels) and trailed observed (centre panels) and reconstructed (right panels) spectra of the H$\alpha$ and H$\beta$ emission lines.}
\label{figure:Dop1a}
\end{figure*}

\begin{figure*}
\centering
 \subfigure[HeII 4686 \AA{} line (2013 October 16--22)]{
   \includegraphics[width = 7.50cm ]{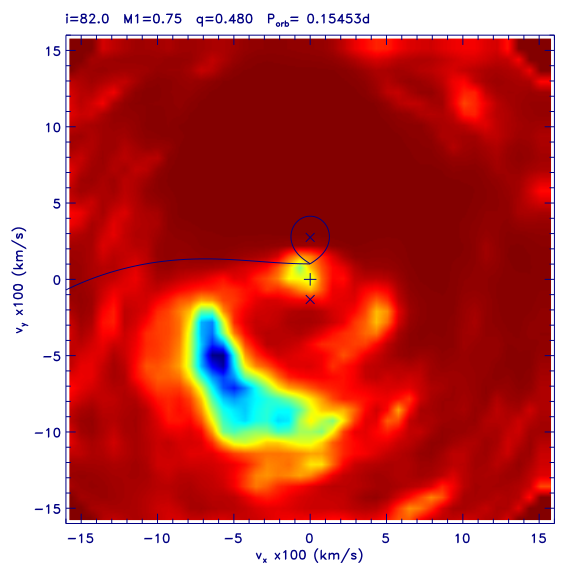}
   \hspace{0.5cm}
   \includegraphics[width = 4.0cm, height = 7.50cm]{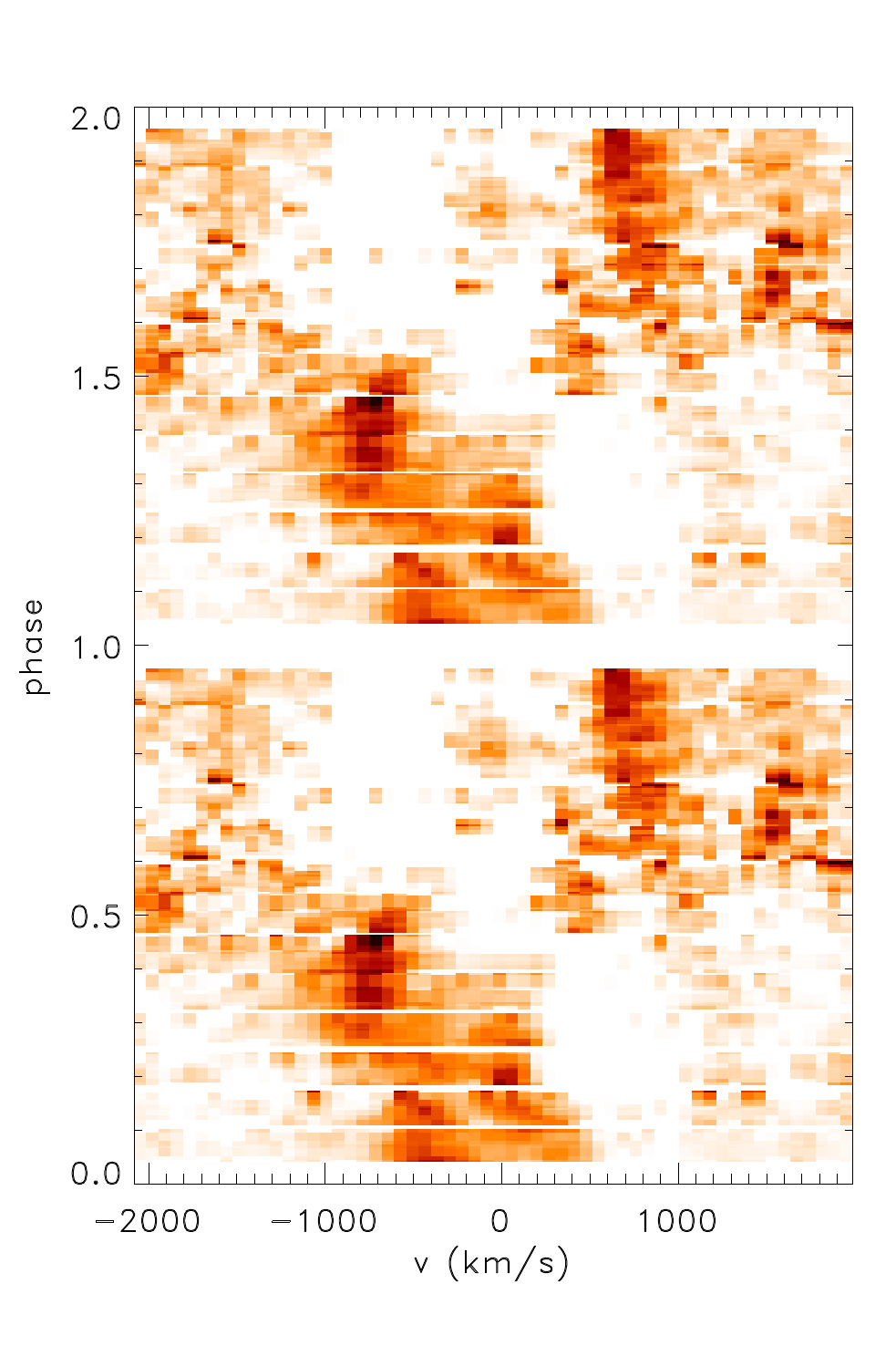}
   \includegraphics[width = 4.0cm, height = 7.50cm]{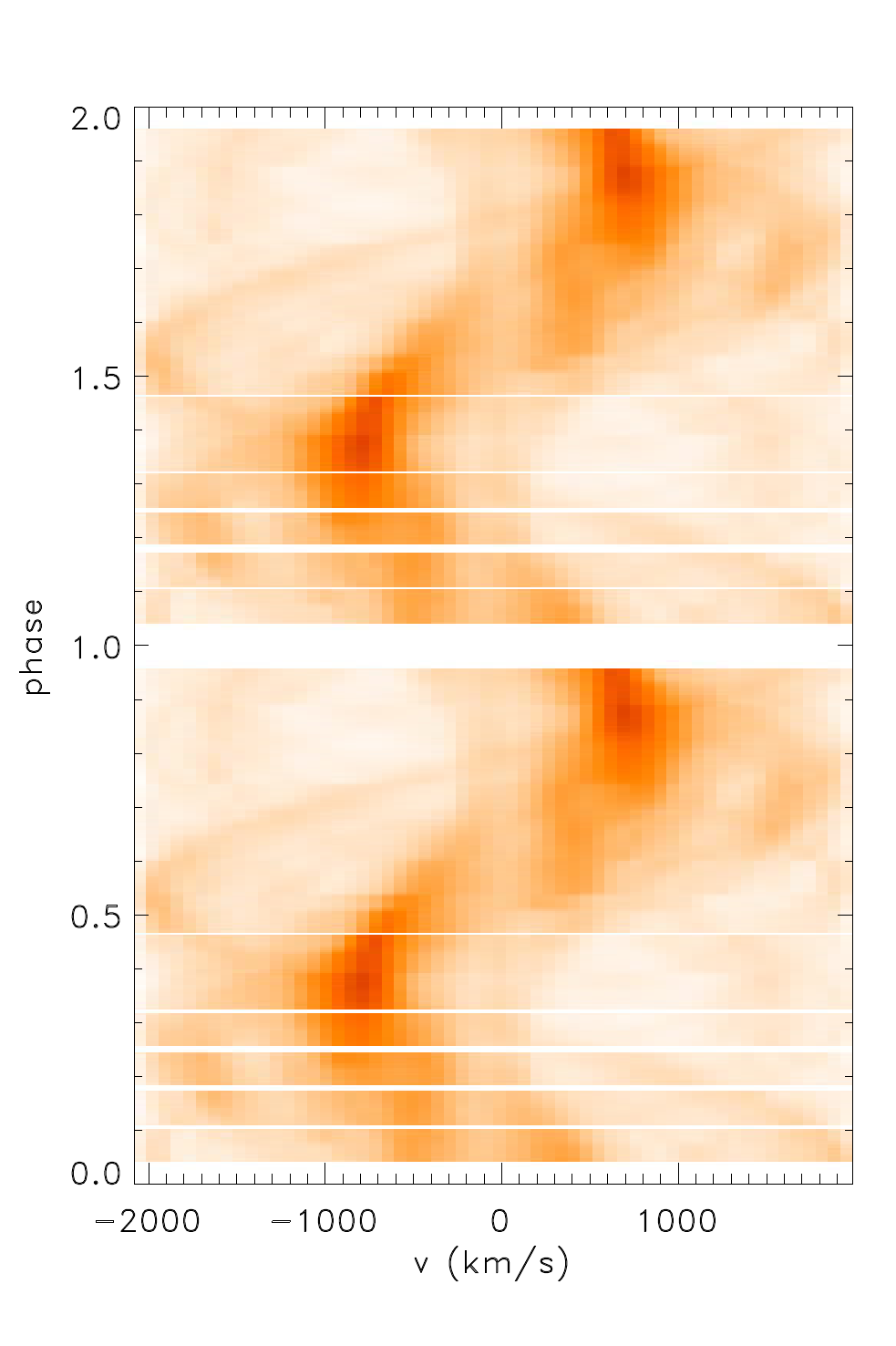}
   \label{figure:dopHeII1013}
   }
 \subfigure[H$\beta$ line (2013 October 16--22)]{
   \includegraphics[width = 7.50cm ]{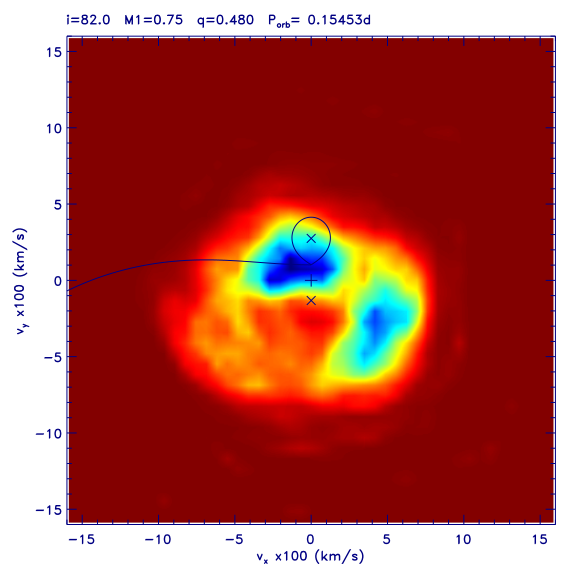}
   \hspace{0.5cm}
   \includegraphics[width = 4.0cm, height = 7.50cm]{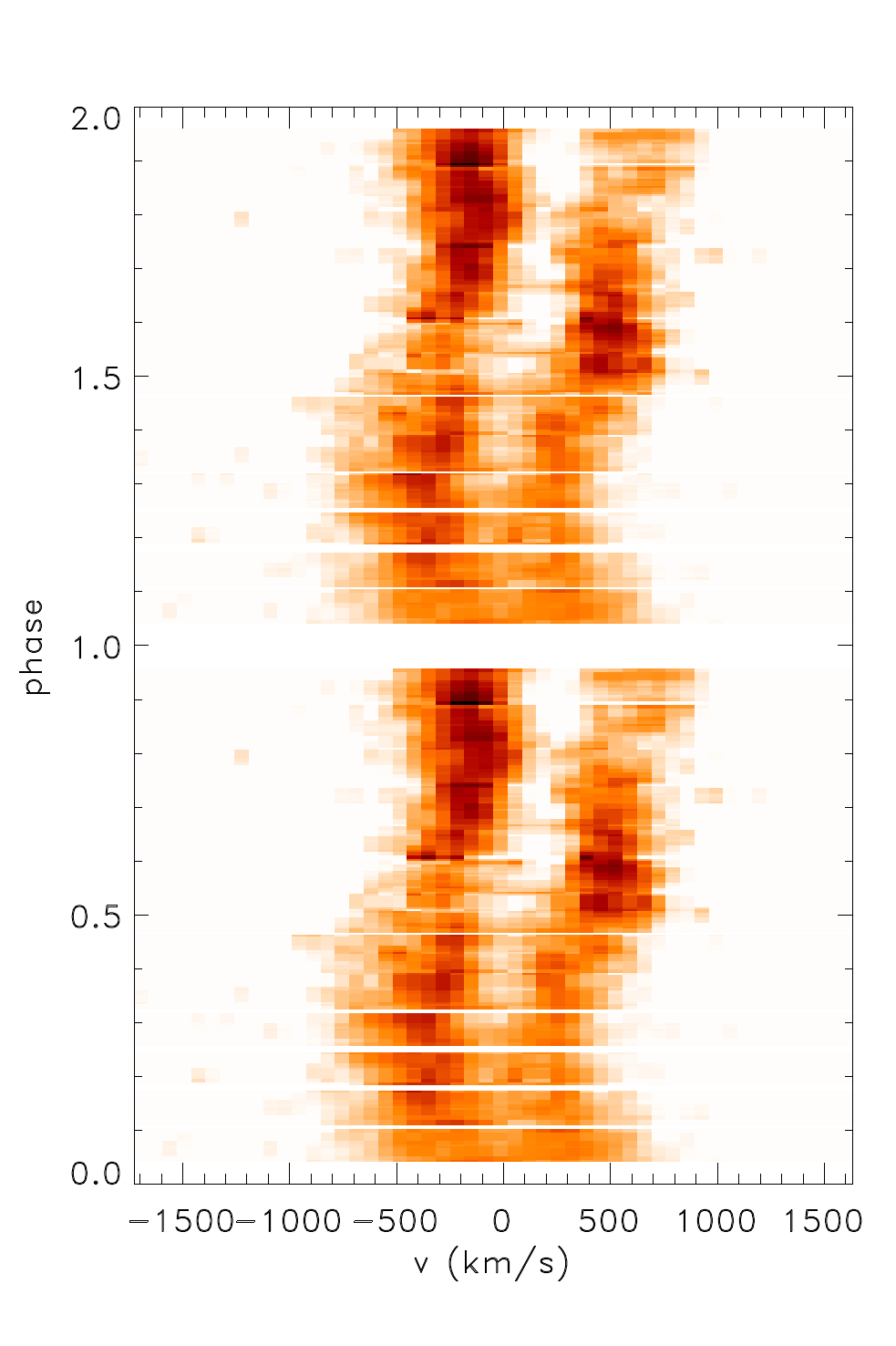}
   \includegraphics[width = 4.0cm, height = 7.50cm]{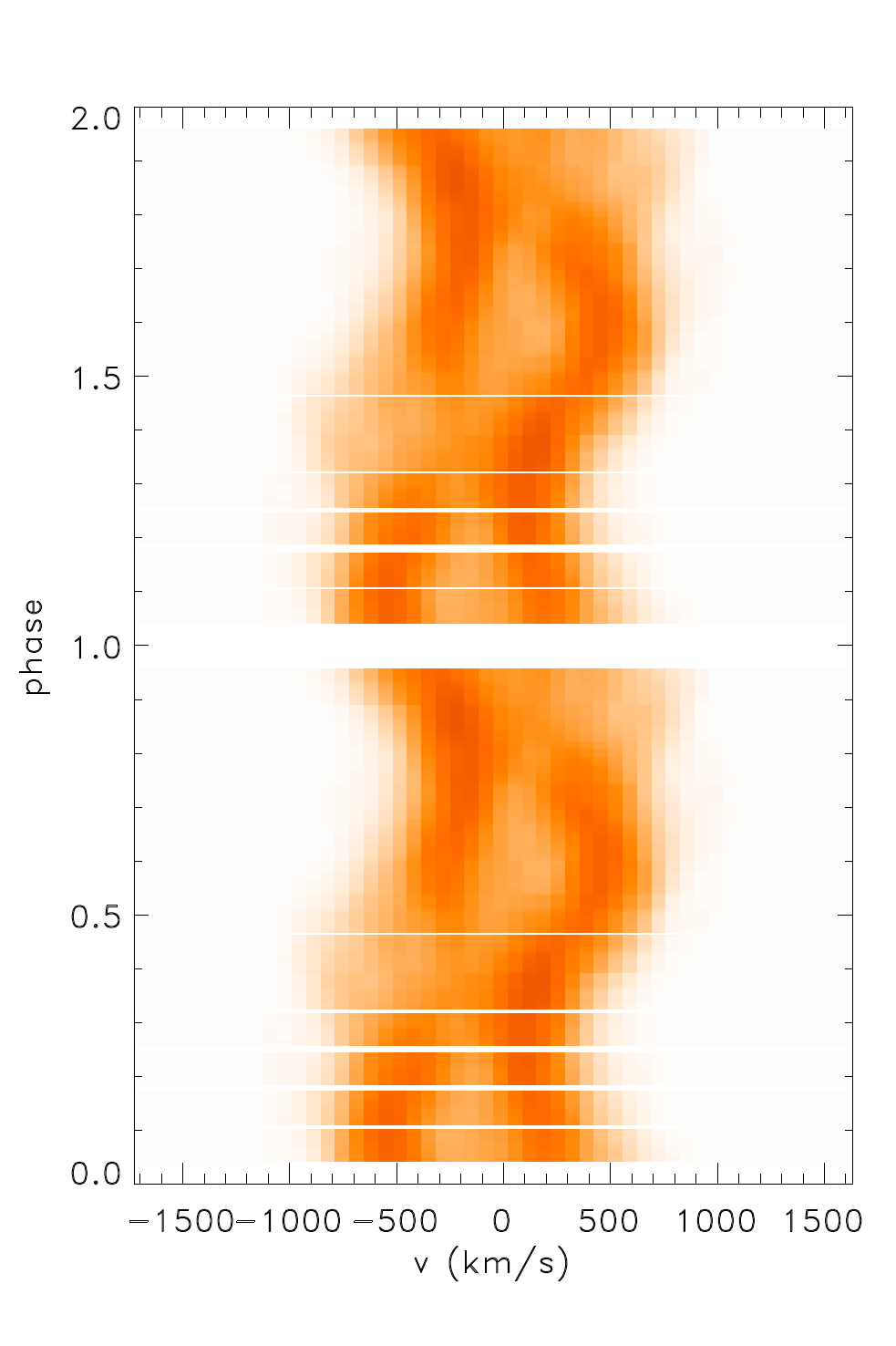}
   \label{figure:dopHbeta1013}
   }
   \caption{Doppler maps (left panels) and trailed observed (centre panels) and reconstructed (right panels) spectra of HeII 4686 \AA{} and H$\beta$ emission lines.}
\label{figure:Dop2}
\end{figure*}

\begin{figure*}
\centering
 \subfigure[H$\gamma$ line (2013 October 16--22)]{
   \includegraphics[width = 7.50cm ]{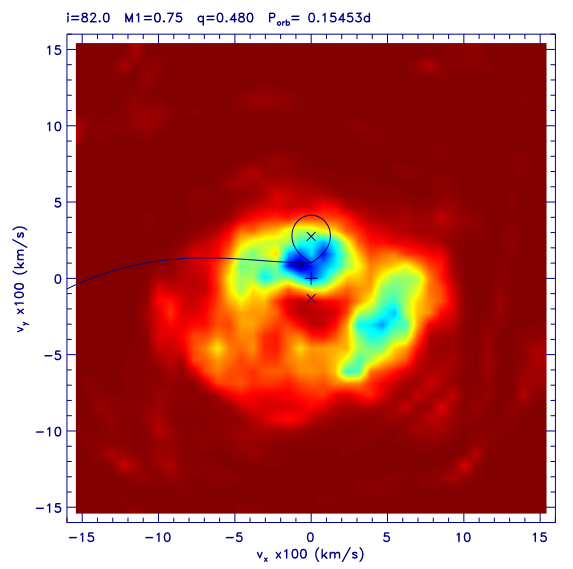}
   \hspace{0.5cm}
   \includegraphics[width = 4.0cm, height = 7.50cm]{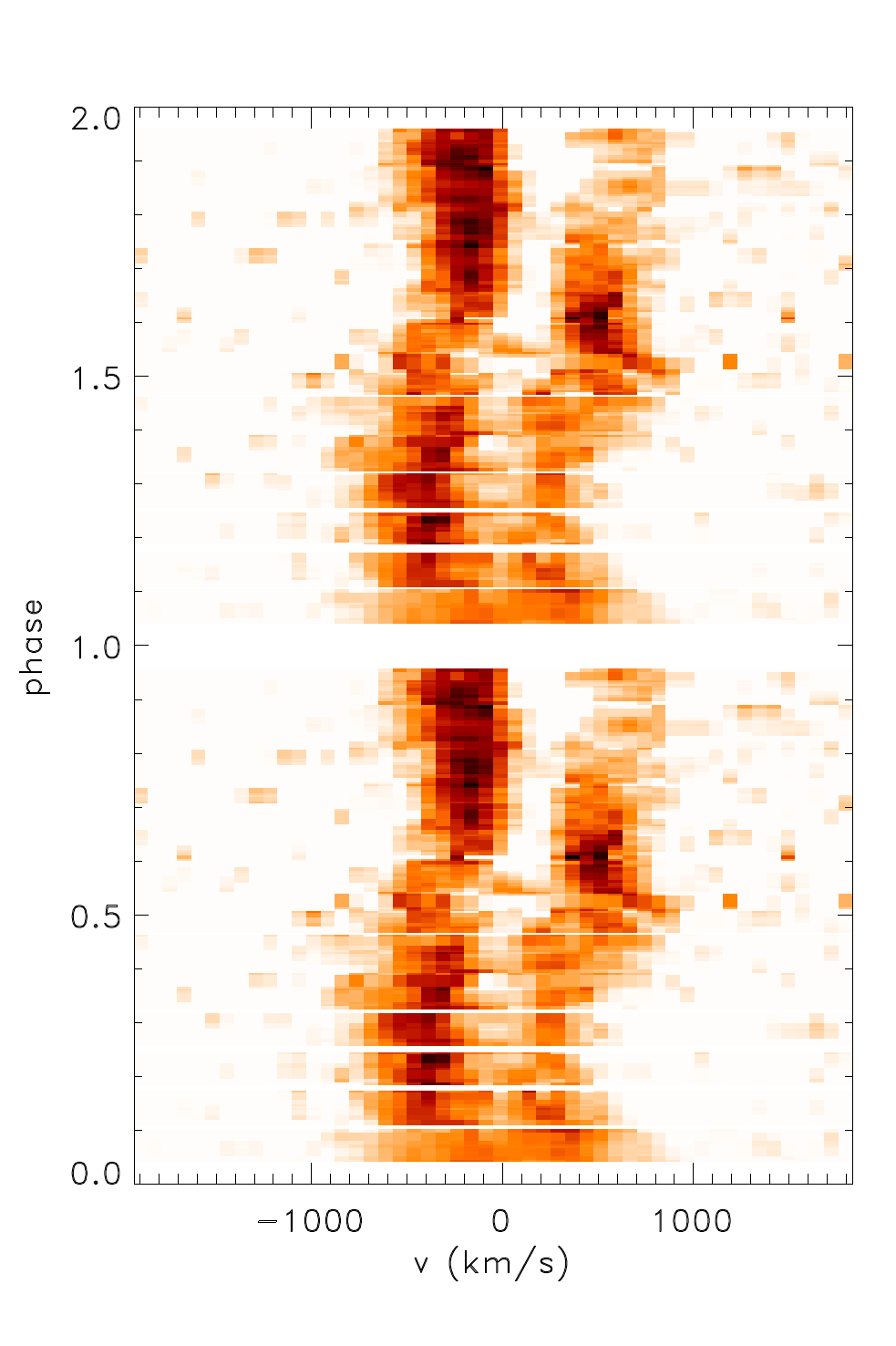}
   \includegraphics[width = 4.0cm, height = 7.50cm]{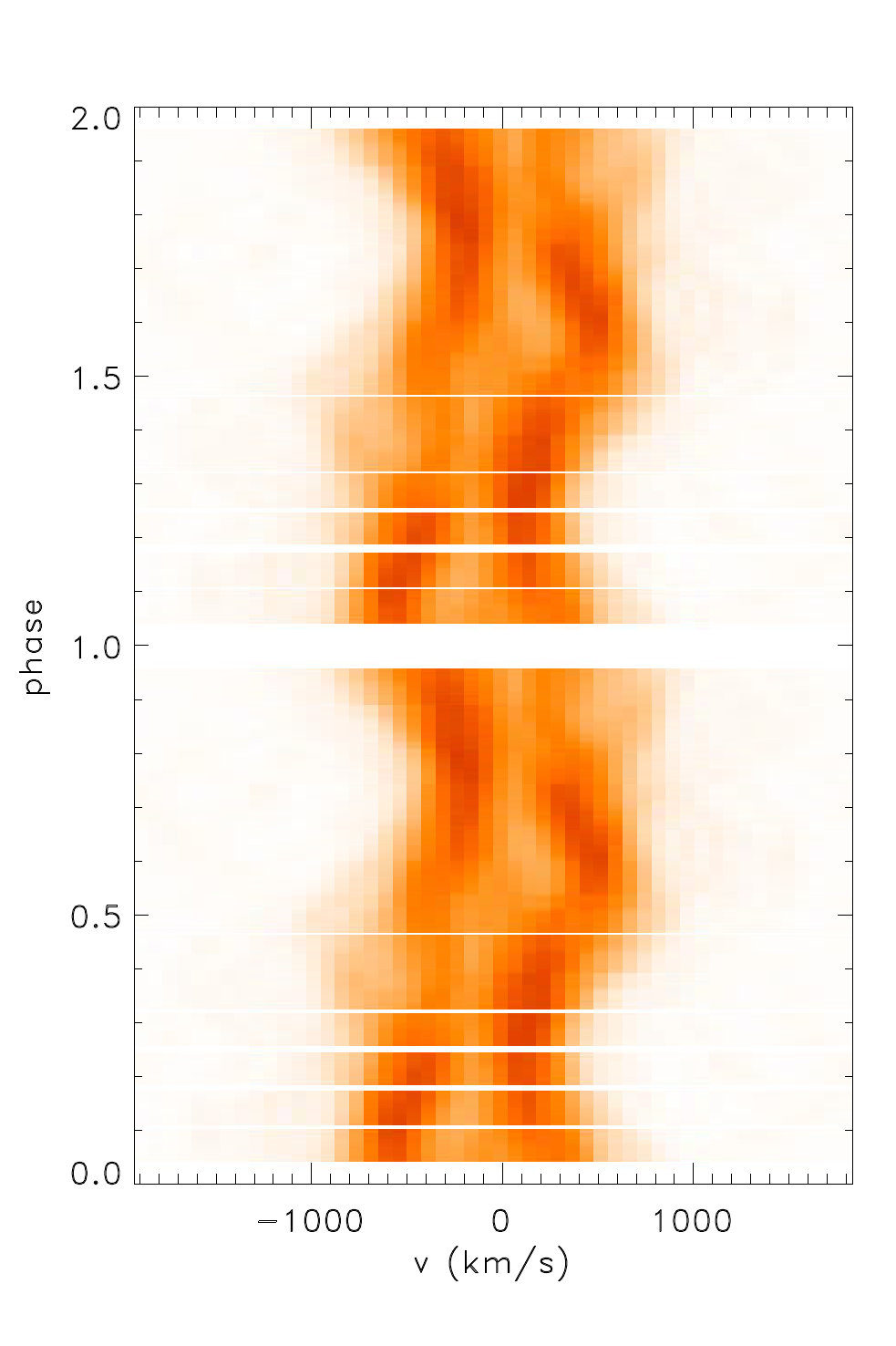}
   \label{figure:dopHgamma1013}
   }
 \subfigure[HeI 4471 \AA{} line (2013 October 16--22)]{
   \includegraphics[width = 7.50cm ]{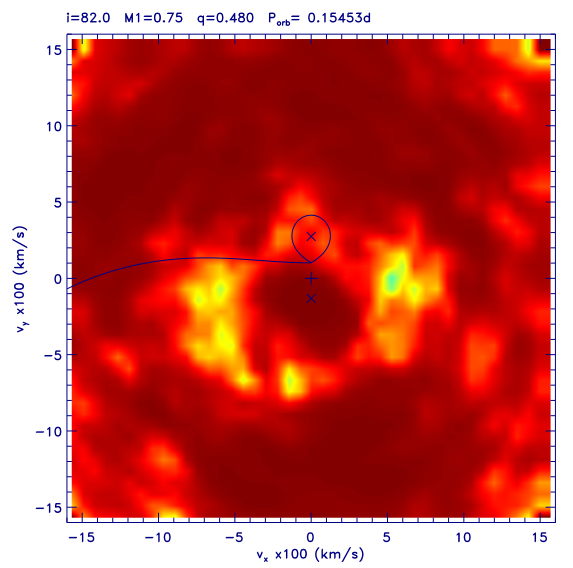}
   \hspace{0.5cm}
   \includegraphics[width = 4.0cm, height = 7.50cm]{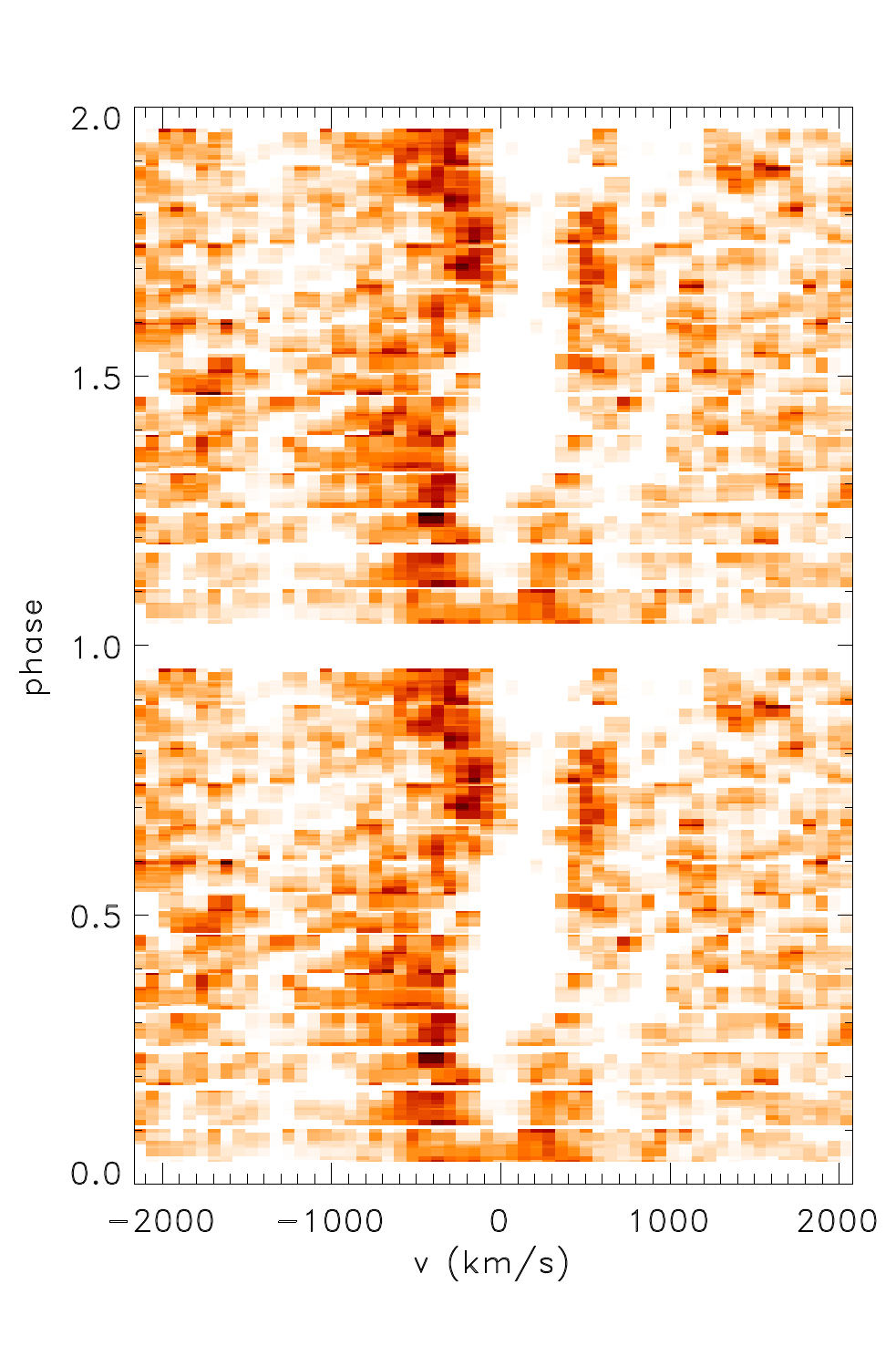}
   \includegraphics[width = 4.0cm, height = 7.50cm]{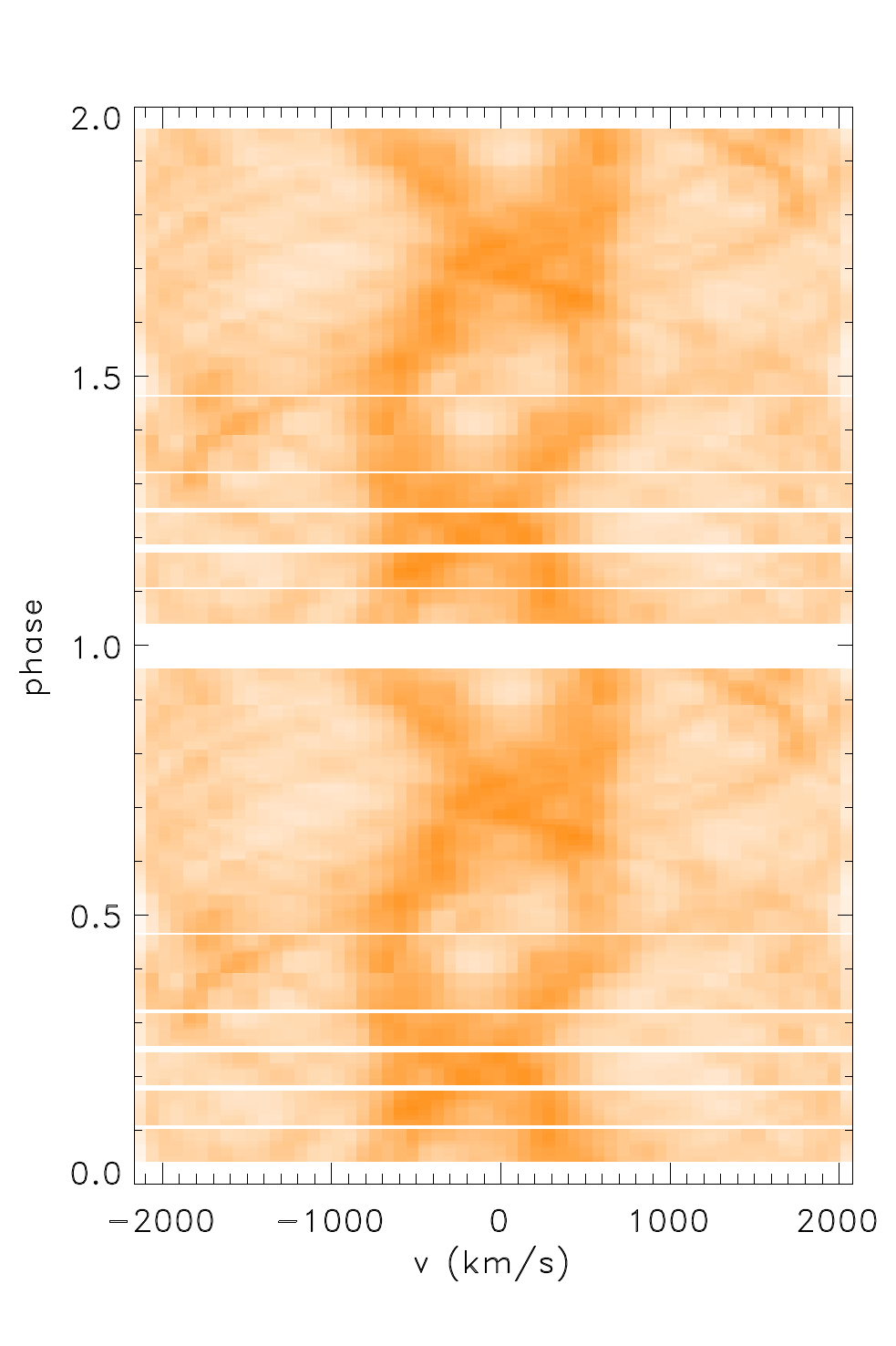}
   \label{figure:dopHeI44711013}
   }
\caption{Doppler maps (left panels) and trailed observed (centre panels) and reconstructed (right panels) spectra of the H$\gamma$ and HeI 4471 \AA{} emission lines.}
\label{figure:Dop2a}
\end{figure*}

We constructed Doppler maps of EC21178-5417 using the fast maximum entropy method (FMEM\footnote{See \url{http://www.mpa-garching.mpg.de/~henk/pub/dopmap/} for more details.}) code developed by \citet{1998astro.ph..6141S} which uses \texttt{IDL}\footnote{The acronym IDL stands for Interactive Data Language and is a trademark of ITT and/or exelis Visual Information Solutions. For further details see \url{http://www.exelisvis.com/ProductsServices/IDL.aspx}.} for plotting. 
We used the mass of the secondary star ($M_2$ = 0.36$M_{\odot}$) and the inclination ($i$) of 82$^{\circ}$ and assumed that the mass of the WD primary ($M_1$) to be $\sim$0.75 $M_{\odot}$ \citep[e.g.][]{1997PhDT........28G,1998MNRAS.301..767S}. 
These gives the mass ratio, $q$ = $M_2/M_1$ = 0.48 and the orbital period ($P\rm_{orb}$ = 0.154527 days) given in Eq. \eqref{eph1}. These parameters were used to compute the Roche-lobe and the stream overlaid on the Doppler maps. 
The location of the primary and secondary is indicated with a \say{$\times$}, whereas that of the centre of mass of the binary is marked with a \say{+}. Similarly, the location of the accretion stream trajectory is indicated by a solid blue line and marks the trajectory followed by the materials moving at stream velocity. 
The Roche-lobe of the secondary star is also indicated. The Doppler maps of EC21178-5417 are shown in Figs \ref{figure:Dop1} to \ref{figure:Dop2a} for the 2011 and 2013 observations. 
For the 2011 observations, the trailed spectra used for the Doppler maps were re-binned into 20 phase bins of 0.05 each. The 2013 observations were re-binned into 40 bins of 0.025 each. The velocity resolution were approximately the same for each observations and was around 70 km/s per pixel.   


\smallskip
The left panel of Figs \ref{figure:dopHeII1011} and \ref{figure:dopHeII0911} show the Doppler maps obtained from the HeII 4686 \AA{} lines. The two panels reveal an asymmetric accretion disc. 
Also, there is evidence of a spiral structure which is interpreted as the indication of the presence of the tidally-induced shock waves in the accretion disc such as that observed during outbursts in the DN IP Peg \citep{1997MNRAS.290L..28S,1999MNRAS.306..348H,1999MNRAS.307...99S,2000MNRAS.313..454M}. 
There is a strong two-armed disc asymmetry visible, especially in Fig. \ref{figure:dopHeII1011}, which is reminiscent of the two-armed spiral structure in the accretion disc. The emission in Fig. \ref{figure:dopHeII0911} is concentrated in the first and third quadrant, and result from the effect of spiral density waves in the accretion disc. 
The non-equal emission of the two sides of the asymmetry lends some support to the models of \citet{2001AcA....51..295S}. 
But there is emission from the region between the WD and the secondary in Fig. \ref{figure:dopHeII1011} which makes the Doppler map difficult to interpret. This feature is seen in the Doppler map from the 2011 October observations. 
We noted that the HeII 4686 \AA{} phase binned spectra from this date (Figure 4.9 of \cite{2013.....Z}) showed multiple components and we therefore associate the emission between the WD and the secondary star to one of the components. 

Similarly, Fig. \ref{figure:dopHeII1013}, shows the Doppler map obtained after subtracting the average spectrum from the individual spectra obtained in 2013 October 16--22. The main reason for doing this was to recover the spiral structure like those shown in Fig. \ref{figure:Dop1} for the same line. 
The original Doppler map of HeII 4686 \AA{} from 2013 October observations showed a blob centered on the WD and it was difficult to interpret. 
It is clear from Fig. \ref{figure:dopHeII1013} that the Doppler map is elongated and contain a spiral structure in the third quadrant. The second arm of the spiral structure is weak, i.e. at a noise level. 

\smallskip
Doppler maps of the Balmer lines: left panel of Figs \ref{figure:dopHalpha0911} and \ref{figure:dopHbeta1011}, and left panel of Figs \ref{figure:dopHbeta1013} and \ref{figure:dopHgamma1013}, also suggest an asymmetric accretion disc similar to that of HeII 4686 \AA{} line but with less defined spiral structure. 
In general, the Doppler maps of the Balmer lines reveal a more circular accretion disc. 
The Doppler map of the H$\alpha$ line (Fig. \ref{figure:dopHalpha0911}) shows no evidence of the emission associated with irradiated secondary, whereas that of H$\beta$ (Fig. \ref{figure:dopHbeta1011} and \ref{figure:dopHbeta1013}) and H$\gamma$ (Fig. \ref{figure:dopHgamma1013}) lines suggest that there could be emission from the secondary star. 
But, the reality of this feature in the Doppler map of H$\beta$ and H$\gamma$ is 
questionable (see section \ref{sec:aa}). 
A possible reason for the absence of the distinct spiral structure in the Doppler map of the Balmer 
lines is that these lines are more strongly affected by the additional absorption component in the lines, making the 
Doppler maps difficult to interpret. Further, no bright spot emission is seen in the Doppler maps of the Balmer lines. 
Figure \ref{figure:dopHeI44711013} shows the Doppler map of HeI 4471 \AA{} which is consistent with that of HeII 4686 \AA{} and the Balmer lines. 
However, most of the features are not clearly visible. As expected, no emission from either the 
hot spot or the secondary star is present on the Doppler map of HeI 4471 \AA{}.


\section{Discussion and Conclusion}
\label{sec:Discussion}

The spectroscopic results presented in Section \ref{sec:SpectraAnalysis} and \ref{sec:Doppler} followed from nine nights worth of observations of EC21178-5417 covering at 
least one orbital cycle of the binary system per night. Analysis of the spectroscopic results of EC21178-5417 reveal that 
it is dominated by double-peaked emission lines from a highly inclined accretion disc. 
The strength of the HeII 4686 \AA{} for EC21178-5417 is similar and/or comparable to that seen during the outbursts of 
IP Peg \citep{1990ApJ...349..593M}. The ratio of HeII 4686 \AA{} to H$\beta$ is greater than unity and the presence 
of HeII 5411 \AA{} suggests that these two lines originate from the region which has higher than normal level of 
ionization, e.g. SW Sex \citep{2001A&A...368..183G}. Further, the presence of HeII at 4200 \AA{} and CII at 
4267 \AA{} which are seen during dwarf novae outbursts suggest that EC21178-5417 is characterized by high mass-transfer 
rate.  

\smallskip
In the last paragraphs of Section \ref{sec:spectra1} we compared the average spectrum of EC21178-5417 taken at different 
observational dates. It shows different spectral characteristics at different dates: from pure absorption in the Balmer 
lines (2011 September) through mixed absorption and emission in the Balmer and HeI lines (2011 October) to pure line 
emission (2002 September and 2013 October). This change in spectral features from emission to absorption is attributed 
to variability of the mass transfer rate from the secondary star to the WD, or, at times EC21178-5417 is dominated by an 
additional absorption component associated with the bright spot. Recently, UX UMa was reported to show similar 
variations on a yet undetermined time scale \citep{2011MNRAS.410..963N}. 
We have seen that the bright spot appears strong at 0.75 $\leq \phi \leq$ 0.95 (Section \ref{sec:bs}), and the 
higher Balmer lines change to absorption when the bright spot is visible. This short lived absorption comes from a 
vertically extended region above the bright spot \citep{2001A&A...368..183G}. Based on these characteristics alone we 
may conclude that EC21178-5417 is a member of either the RW Tri or UX UMa class of NLs since the double-peaked emission 
lines rule out the SW Sex membership.

\smallskip 

The trailed spectra of HeII 4686 \AA{} lines show two peaks moving in anti-phase, whereas that of the Balmer lines show two peaks moving in-phase with each other (see Section \ref{sec:aa}). The anti-phase direction between the two peaks result from a spiral structure or wave in the accretion disc. 
The observed and reconstructed trailed spectra of EC21178-5417 were consistent between the three observations, except for the H$\beta$ and H$\gamma$ line which suggested a third component with low velocity amplitude -- that is associated with emission from the secondary star. 
This is attributed to one of the Doppler tomography axioms which state that the intensity of the line emission remains constant throughout the orbital phase. The Doppler tomography code has added a third component to compensate for this varying line emissions. 

\smallskip
Doppler maps of HeII 4686 \AA{} for EC21178-5417 (Section \ref{sec:dop}) reveal the presence of an asymmetric and highly inclined accretion disc with spiral structures. The asymmetry and non-equal intensity of the two regions of emissions is similar to that observed in some DN during outbursts, e.g. IP Peg \citep[eg.][]{1999MNRAS.307...99S,2000MNRAS.313..454M}, SS Cyg \citep{1996MNRAS.281..626S,2012A&A...538A..94K}, EX Dra \citep{2000A&A...354..579J,2000A&A...356L..33J}, U Gem \citep[eg.][]{2001ApJ...551L..89G} and WZ Sge \citep{2002PASJ...54L...7B,2004AN....325..185S}. 
In NLs, spiral structure has been seen in the Doppler maps of three other systems, e.g. V347 Pup \citep{1998MNRAS.299..545S,2005MNRAS.357..881T}, V3885 Sgr \citep{2005MNRAS.363..285H}, and UX UMa \citep{2011MNRAS.410..963N}. 
This makes EC21178-5417 the fourth NL to show spiral structure in its Doppler map. But in previous three systems, spiral structures were seen in the Balmer and HeI lines. 
The spiral structures found in EC21178-5417 are similar to those found in the Doppler maps of IP Peg \citep[e.g.][]{1997MNRAS.290L..28S,1999MNRAS.306..348H} and EX Dra \citep{2000A&A...356L..33J} during outbursts. 
More specifically, the spiral arms in the HeII 4686 \AA{} line of EX Dra do not follow a circle centred on the WD but are rather elongated along the ($V_x,V_y$) axis, the same hold for EC21178-5417 (see Figs \ref{figure:Dop1} and \ref{figure:Dop2}). 
The Doppler map of the Balmer lines reveal a more circular accretion disc, and are harder to interpret due to the time-variable absorption component. 
There is no emission from the secondary star in the Doppler maps of the Balmer lines and the blob/spot in the vicinity of the secondary is either an extension of the disc or the spiral structure. Also, there is no evidence of emission from near the WD in the Doppler maps of EC21178-5417. 

\smallskip
In summary, we presented spectroscopic analysis and Doppler tomography of the eclipsing NL EC21178-5417. 
It has shown persistent spiral structure in its Doppler maps on three occasions and is now one of the four NLs binary systems after V347 Pup, V3885 Sgr and UX UMa to show spiral density waves in its tomograms. 
A follow-up study on the night-to-night behavior of the spiral structures in EC21178-5417 has been conducted by \citet{2020MNRAS.491..344R}.

\section*{Acknowledgments}
We would like to thank the anonymous referee whose comments were helpful and improved our manuscript. 
ZNK acknowledges the postgraduate bursary from the National Research Foundation (NRF) of South Africa through National Astrophysics and Space Science Programme (NASSP) and the internship from the Department of Science and Innovation of South Africa (Ref. N0: S703-8000-887) leading to these results. PAW and BW acknowledge financial support from the NRF and the University of Cape Town (UCT). DK acknowledges the University of the Western Cape (UWC) and the NRF for financial support.

\bibliographystyle{mnras}
\bibliography{references}

\bsp	
\label{lastpage}
\end{document}